\documentclass[%
 reprint,
 amsmath,amssymb,
 aps,
 prl,
]{revtex4-2}

\usepackage{graphicx}
\usepackage{dcolumn}
\usepackage{bm}
\usepackage{caption}
\usepackage{xcolor}
\usepackage{ulem}
\usepackage{amsmath,amssymb,amsbsy,amsfonts,latexsym}
\usepackage{hyperref}
\usepackage[caption=false]{subfig}

\makeatletter
\newenvironment{SMbibliography}[1]
  {\section*{References}%
   \list{\@biblabel{\@arabic\c@enumiv}}%
        {\settowidth\labelwidth{\@biblabel{#1}}%
         \leftmargin\labelwidth
         \advance\leftmargin\labelsep
         \usecounter{enumiv}%
         \setlength{\itemsep}{0pt}%
         \setlength{\parsep}{0pt}%
         \setlength{\topsep}{0pt}%
         \let\p@enumiv\@empty
         \renewcommand\theenumiv{\@arabic\c@enumiv}}%
   \sloppy\clubpenalty4000\widowpenalty4000%
   \sfcode`\.=\@m}
  {\def\@noitemerr{\@latex@warning{Empty `SMbibliography' environment}}%
   \endlist}
\makeatother

\newmuskip\pFqmuskip

\newcommand\fH[1]{\sbox0{#1}\dimen0=\ht0 \advance\dimen0 -1ex
	\sbox2{\'{}}\sbox2{\raise\dimen0\box2}%
	{\ooalign{\hidewidth\kern.1em\copy2\kern-.5\wd2\box2\hidewidth\cr\box0\crcr}}}

\newcommand*\pFq[6][8]{%
	\begingroup
	\pFqmuskip=#1mu\relax
	\mathchardef\normalcomma=\mathcode`,
	\mathcode`\,=\string"8000
	\begingroup\lccode`\~=`\,
	\lowercase{\endgroup\let~}\pFqcomma
	{}_{#2}F_{#3}{\left[\genfrac..{0pt}{}{#4}{#5};#6\right]}%
	\endgroup
}
\newcommand{\pFqcomma}{{\normalcomma}\mskip\pFqmuskip}

\newcommand{\bq}{\begin{equation}}
\newcommand{\eq}{\end{equation}}
\newcommand{\bqa}{\begin{eqnarray}}
\newcommand{\eqa}{\end{eqnarray}}
\newcommand{\nn}{\nonumber \\}
\def\be     {\begin{equation}}
\def\ee     {\end{equation}}
\def\bea        {\begin{eqnarray}}
\def\eea        {\end{eqnarray}}
\def\bnn    {\begin{eqnarray*}}
\def\enn    {\end{eqnarray*}}

\begin{document}

\preprint{APS/123-QED}

\title{Finite-momentum inter-orbital superconductivity driven by chiral charge-density-wave quantum criticality beyond the BCS regime}

\author{Jin Mo Bok}%
\email{jinmobok@postech.ac.kr}
\author{B. J. Kim}
\author{Ki-Seok Kim}%
\email{tkfkd@postech.ac.kr}
\affiliation{Department of Physics, Pohang University of Science and Technology, Pohang 37673, South Korea}

\date{\today}

\begin{abstract}
Superconductivity emerging near charge-density-wave (CDW) quantum critical points often defies a conventional BCS description, particularly in multi-orbital systems with small and orbitally distinct Fermi surfaces. In TiSe$_2$, superconductivity appears under pressure near the suppression of a chiral CDW, yet its microscopic origin has remained unresolved.
Here we show that the chiral CDW quantum criticality in TiSe$_2$ originates from a fluctuation-induced intertwining of charge-order and phonon modes that are symmetry incompatible at the Brillouin-zone center but become mixable at the CDW ordering wave vector.
This resolution of symmetry frustration enables a single continuous chiral CDW transition and strongly enhances collective fluctuations near criticality.
We demonstrate that these critical chiral CDW fluctuations drive a finite-center-of-mass-momentum inter-orbital pairing instability fundamentally different from BCS superconductivity. Because electrons near the $\Gamma$ and $L$ points occupy small $p$- and $d$-orbital Fermi pockets connected only by the CDW ordering vector, the inter-orbital pair susceptibility does not develop a Cooper logarithm. As a result, superconductivity is governed by an interaction-driven pairing mechanism rather than by the density of states.
Using a symmetry-constrained low-energy theory and a random-phase-approximation analysis, we show that the fluctuation-enhanced pairing interaction is maximized near the chiral CDW quantum critical point, giving rise to a dome-shaped superconducting phase. A group-theoretical analysis further identifies an orbital-selective $s$-wave pairing symmetry as the most likely superconducting state.
\end{abstract}

\maketitle

\begin{figure*}
\centering
\hfill
\subfloat[]{\includegraphics[width=6.5cm]{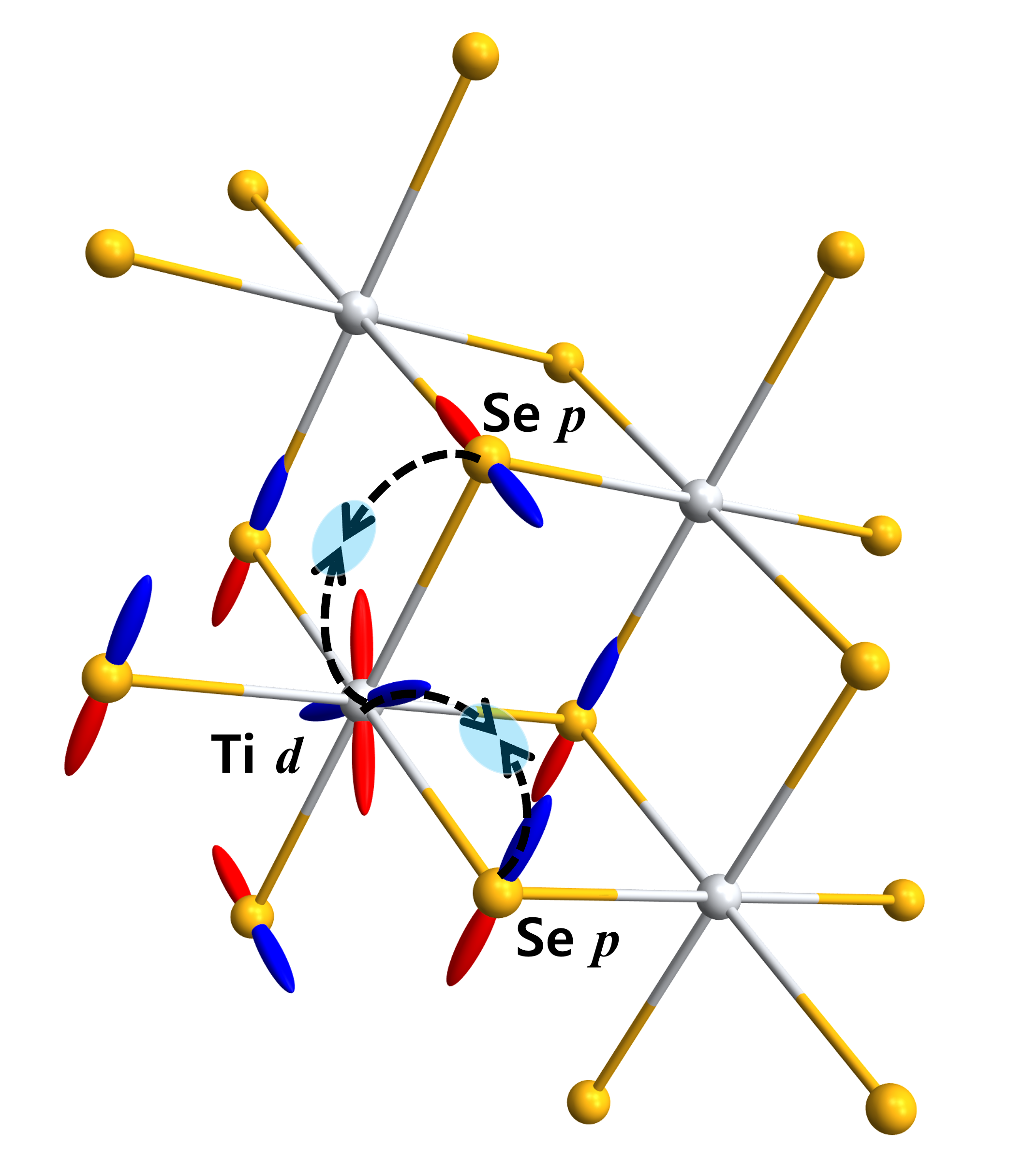}}
\hfill
\subfloat[]{\includegraphics[width=9.5cm]{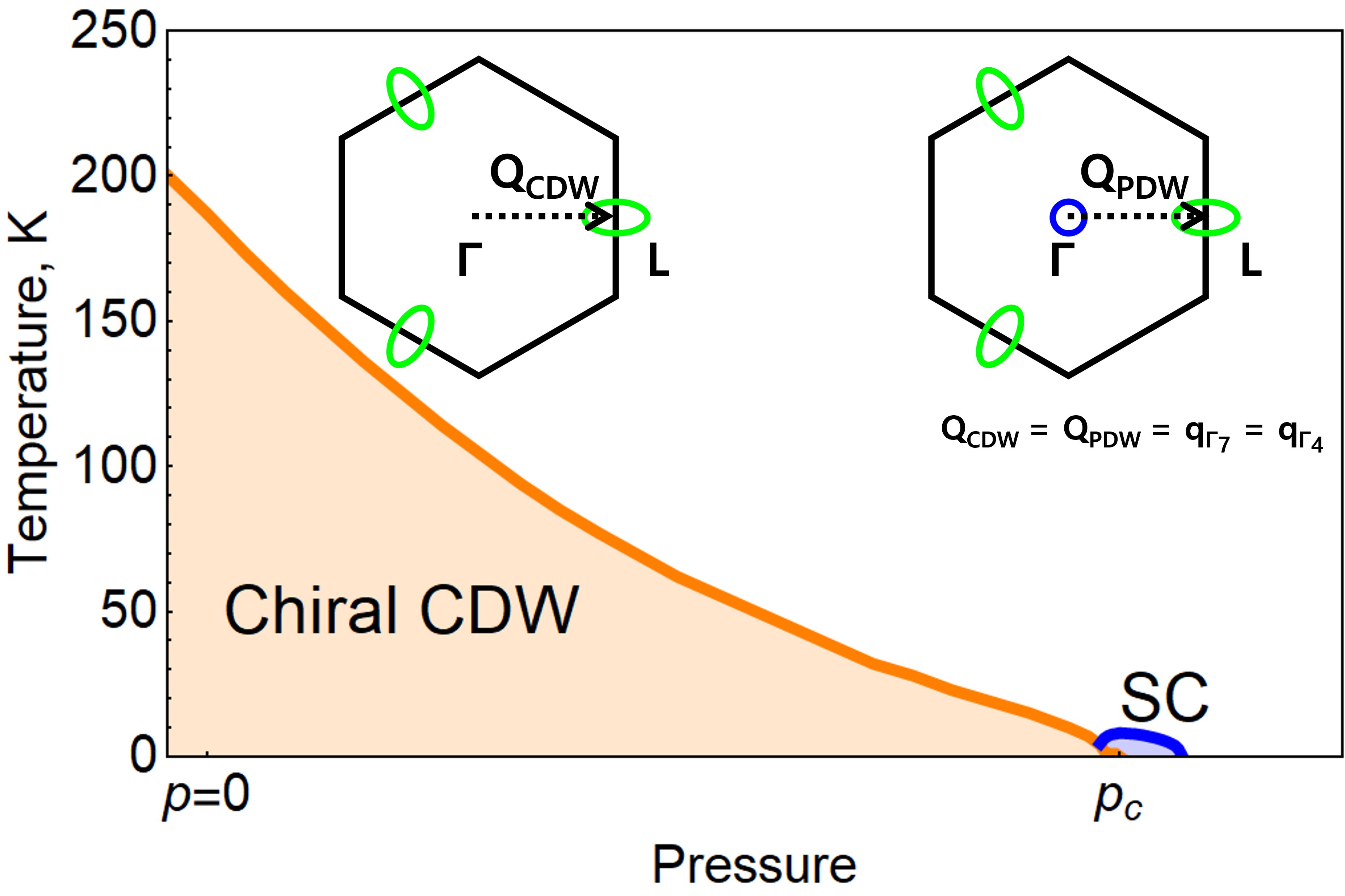}}
\hfill
\captionsetup{justification=raggedright}
\caption{Left: Schematic representation of the crystal structure and inter-orbital Cooper pairing of electrons from $p$ and $d$ orbitals.
In momentum space, $p$-orbitals and $d$-orbitals reside near $\Gamma$ and $L$ point, respectively.
Right: The phase diagram of chiral CDW and SC based on many-body calculations in the RPA fashion.
$T_\text{CDW}$ and $T_c$ are given by the instability condition of Eq. (\ref{eq:phonon}) and Eq. (\ref{eq:cooper}), respectively.
The inset shows the Lifshitz transition of the hole Fermi surface (blue) at the $\Gamma$ point, occurring near the critical pressure.
Cooper pairs induced by quantum critical chiral CDW fluctuations have a total momentum of $\mathbf{Q}_\text{PDW}=\mathbf{Q}_\text{CDW}$ ($=\mathbf{q}_{\Gamma_7}=\mathbf{q}_{\Gamma_4}$).
}
\label{fig:main}
\end{figure*}

\section{Introduction}
Charge-density-wave (CDW) order is often viewed as a simple collective state characterized by a periodic modulation of electronic charge density coupled to lattice distortions. This conventional picture, however, can break down in multi-orbital van der Waals materials, where orbital degrees of freedom intertwine with space-group symmetries and give rise to particle–hole composite order parameters transforming under nontrivial irreducible representations, as demonstrated in recent studies of orbital-driven and axial Higgs modes in layered charge-density-wave systems.\cite{Axiality_RTe3_Exp,Axiality_RTe3_Th,Wulferding2025} As a result, charge ordering in such systems may involve additional symmetry breaking beyond that expected in single-orbital settings.
A prominent example is TiSe$_2$, which hosts an unconventional chiral CDW state. Kim \textit{et al}.\cite{Kim2024} proposed that symmetry frustration between charge-ordering and phonon modes belonging to incompatible irreducible representations can be resolved through fluctuation-driven symmetry breaking, leading to a “complete” lifting of the underlying symmetry constraints and stabilizing the chiral CDW phase.\cite{Ishioka2010,Wezel2011,Xu2020} This state develops near 200~K\cite{Holt2001} and exhibits strong sensitivity to non-thermal control parameters.

The chiral CDW phase in TiSe$_2$ can be continuously suppressed by non-thermal tuning parameters such as external pressure,\cite{Joe2014,Kusmartseva2009,Kitou2019,Xu2021,Hinlopen2024} chemical intercalation,\cite{Morosan2006,Zhao2007,Wu2007,Li2007,Kogar2017prl,Liao2021,Piatti2023} and electrostatic gating.\cite{Li2016} Near the critical point where the chiral CDW collapses, superconductivity emerges and forms a dome-shaped phase with a maximum $T_c$ of approximately $2$–$4$~K,\cite{Li2016,Liao2021,Kusmartseva2009,Morosan2006,Hinlopen2024} reminiscent of other strongly correlated superconductors.
A growing body of experimental evidence indicates that this superconducting state deviates from a conventional BCS description. Superconductivity develops near the commensurate–incommensurate CDW boundary,\cite{Joe2014,Li2016,Kogar2017prl} coincides with a pressure-induced Lifshitz transition,\cite{Hinlopen2024} and is accompanied by anomalous metallic behavior and signatures of multi-gap superconductivity.\cite{Liao2021,Li2019,Zaberchik2010,Piatti2025}
Moreover, scaling behavior of the CDW amplitude mode,\cite{Barath2008} together with phonon softening and enhanced electronic fluctuations near the CDW endpoint,\cite{Snow2003,Lioi2016} points to a chiral CDW quantum critical regime that is intimately linked to the emergence of superconductivity and calls for a unified microscopic understanding.
Alternative interpretations include superconductivity nucleating at incommensurate CDW domain walls,\cite{Joe2014,Li2016,Kogar2017prl} conventional electron--phonon--mediated intraband pairing,\cite{Calandra2011} and renormalization-group analyses of electronically driven pairing in two-dimensional TiSe$_2$\cite{Ganesh2014}; these possibilities will be revisited in the Discussion.

This motivates a closer examination of how the electronic structure evolves across the CDW–superconductivity transition. In pristine TiSe$_2$, small electron pockets reside near the $L$ points, while the $\Gamma$-centered hole band lies below the Fermi level,\cite{Rasch2008,May2011,Chen2015,Campbell2019,Watson2019,Yilmaz2024,Buchberger2025} resulting in poor Fermi-surface nesting and rendering a conventional Peierls-type instability unlikely.\cite{Holt2001,Kogar2017} Instead, the CDW is widely understood to involve interband electronic processes, such as exciton condensation.\cite{Kogar2017}
Because the CDW is tightly linked to the underlying electronic structure,\cite{Rasch2008,Bok2021} its evolution under external tuning becomes particularly consequential when superconductivity emerges. In particular, the reconstruction of the $\Gamma$-centered states as the CDW is suppressed is expected to play a central role in shaping the superconducting instability.\cite{Hinlopen2024} In this sense, TiSe$_2$ provides a concrete example of an electronic-structure–driven route to superconductivity in multi-orbital chiral CDW systems.

Here we demonstrate that pressurized TiSe$_2$ hosts an inter-orbital superconducting instability in which electrons from Se $p$ orbitals near $\Gamma$ and Ti $t_{2g}$ orbitals near $L$ form the dominant pairing channel.
Because these small, orbitally distinct Fermi pockets are connected only by the chiral CDW ordering vector, the resulting Cooper pairs necessarily carry a finite-center-of-mass-momentum $Q = Q_{\mathrm{CDW}}$, realizing an intrinsic finite-momentum pairing instability without magnetic fields.
Crucially, the associated $p$–$d$ pair susceptibility remains nearly temperature independent and does not develop a BCS Cooper logarithm, indicating that superconductivity is governed by an interaction-driven mechanism rather than by the density of states.

Instead, we show that superconductivity in TiSe$_2$ is governed by a fluctuation-enhanced effective interaction that is strongly amplified near the chiral CDW quantum critical point, naturally giving rise to a dome-shaped superconducting phase.
We develop a symmetry-constrained low-energy description of chiral CDW fluctuations and their coupling to the electronic degrees of freedom, and show that the resulting interaction mediated by coupled CDW and phonon modes drives an inter-orbital $\Gamma$–$L$ pairing instability in the absence of a Cooper logarithm.
As a consequence, the superconducting transition is controlled by the interaction strength rather than by the density of states.
A group-theoretical analysis further identifies an orbital-selective $s$-wave pairing symmetry, consistent with experimental observations of nodeless superconductivity in TiSe$_2$.\cite{Li2007,Zaberchik2010}
Taken together, our results establish a coherent microscopic framework in which chiral CDW quantum criticality generically drives finite-momentum, inter-orbital superconductivity in multi-orbital systems.

\section{Results}
\begin{figure}
\includegraphics[width=7.5cm]{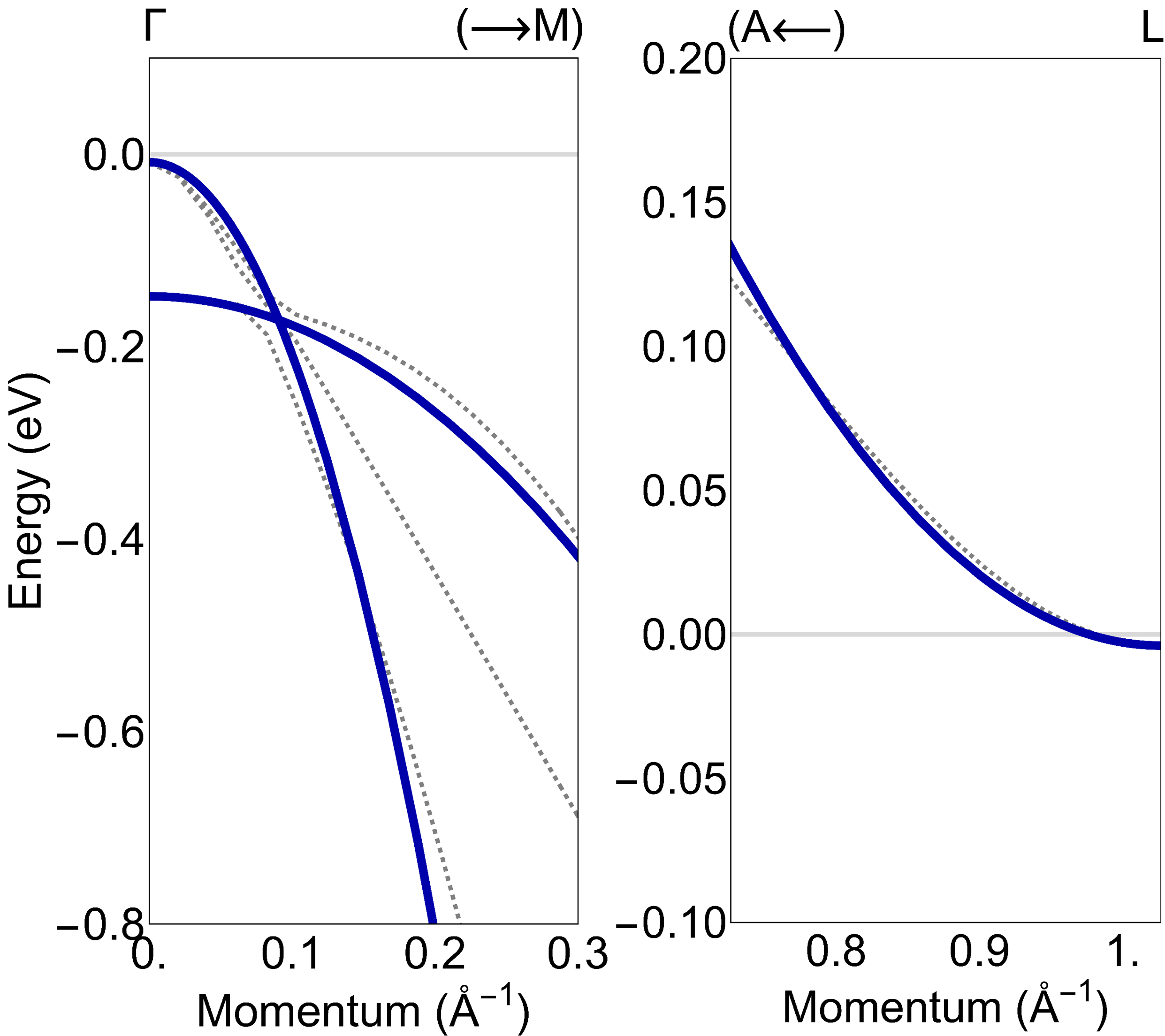}
\captionsetup{justification=raggedright}
\caption{\label{fig:dispersion} The electronic structure near the $\Gamma$ and $L$ point.
The grey dashed line is the tight binding dispersion from Monney $\textit{et al}$\cite{Monney2011dft}.
The blue line is the symmetry-constrained dispersion that we obtain.}
\end{figure}

Pressure tuning provides an ideal route to access the chiral CDW quantum critical point (QCP) in TiSe$_2$ while avoiding doping-induced disorder.\cite{Kusmartseva2009,Joe2014}
In this regime, the low-energy electronic structure is governed by small and orbitally distinct Fermi pockets: Ti $t_{2g}$-derived electron pockets near the three $L$ points and a Se $p$-derived hole band near $\Gamma$,\cite{Kaneko2018} which is driven upward by a pressure-induced Lifshitz transition\cite{Hinlopen2024} from below the Fermi level at ambient pressure.\cite{Rasch2008,May2011,Chen2015,Campbell2019,Watson2019,Yilmaz2024,Buchberger2025}
This experimentally established evolution, evidenced by ARPES and transport measurements,\cite{Kusmartseva2009,Xu2021} defines the minimal kinematic setting in which finite-momentum inter-orbital pairing emerges naturally, since the relevant $\Gamma$ and $L$ states communicate primarily through the CDW ordering vector $Q_{\mathrm{CDW}}$.
A symmetry-constrained low-energy description captures this structure by encoding pressure as an upward shift of the $\Gamma$-centered hole band toward the Fermi level, consistent with Fig.~\ref{fig:dispersion}.\cite{Monney2011dft}

A key ingredient for realizing the chiral CDW through a single continuous transition is an effective coupling between the charge-ordering mode and the phonon mode at the ordering wave vector, even though these modes belong to different irreducible representations. At the Brillouin-zone center, experiments identify the charge modulation with the $\Gamma_4$ sector and the relevant phonon with the $\Gamma_7$ sector,\cite{Kim2024,Ishioka2010} which transform under distinct irreducible representations of the full $D_{3d}$ point group. As a result, a direct linear coupling between these two modes is strictly forbidden at $\Gamma$, since the product representation $\Gamma_4 \otimes \Gamma_7$ does not contain the totally symmetric representation required for a scalar invariant.

This symmetry constraint, however, is fundamentally altered at finite ordering wave vector $q = Q_{CDW}$. Away from the Brillouin-zone center, the relevant symmetry operations are no longer those of the full point group at $\Gamma$, but are instead governed by the little group of $Q_{CDW}$, which contains a reduced set of symmetry elements. Under this reduced symmetry, the $\Gamma_4$ and $\Gamma_7$ modes---defined as irreducible at $\Gamma$---no longer constitute good symmetry labels and are effectively reclassified into compatible representations of the little group. Consequently, their product can acquire a scalar component, rendering a linear mixing between the charge-ordering and phonon sectors symmetry allowed at $q = Q_{CDW}$.

Microscopically, this coupling is dynamically generated by particle-hole fluctuations between the small $\Gamma$-centered hole pocket and the $L$-centered electron pockets, which naturally carry momentum $Q_{CDW}$.
Interband $\Gamma$--$L$ fluctuations dynamically mix the $\Gamma_4$ CDW field with the $\Gamma_7$ sector at $q = Q_{CDW}$, giving rise to the fluctuation-induced coupling
\begin{equation}\label{eq:comver}
\phi_{\Gamma_4}^{\alpha\dagger}(\mathbf{q})\,\Pi^{\alpha\beta}(\mathbf{q})
\big[a_{\Gamma_7}^{\beta}(\mathbf{q})+\phi_{\Gamma_7}^{\beta}(\mathbf{q})\big].
\end{equation}
In Eq.~(\ref{eq:comver}), the polarization matrix $\Pi^{\alpha\beta}(\mathbf q)$ denotes the interband particle--hole bubble connecting the $\Gamma$- and $L$-centered electronic states.
In a fully microscopic description, this object is dressed by momentum- and orbital-dependent form factors arising from the electronic vertices that couple to the $\Gamma_4$ and $\Gamma_7$ sectors.
Such vertex structures generally reduce the overall magnitude of the polarization amplitude due to their internal orbital texture, while leaving the symmetry-allowed mixing between the two sectors intact.
In the present work, we do not include these form-factor effects explicitly in $\Pi$, and instead retain only the symmetry-constrained matrix structure that captures the essential mixing mechanism.
The associated reduction of the effective coupling strength should therefore be understood as being absorbed into the subsequent RPA resummation of the collective-mode propagator discussed below.
This fluctuation-induced mixing resolves the symmetry frustration between the $\Gamma_4$ and $\Gamma_7$ sectors and enables a single continuous transition in which the primary CDW order condenses at $T_{CDW}$, consistent with experimental observations.\cite{Kim2024}

As a consequence of the fluctuation-induced coupling between the $\Gamma_4$ and $\Gamma_7$ sectors, the phonon dynamics at $q = Q_{CDW}$ is strongly renormalized.
The resulting phonon propagator incorporates the full effect of electronic collective-mode fluctuations and is given by
\begin{eqnarray}\label{eq:phonon}
&& D^{\alpha\beta}_{\text{RPA}}(i \Omega_{m},\bm{q}) = \frac{1}{2} \Big\langle a_{\Gamma_{7}}^{\alpha}(i \Omega_{m},\bm{q}) a_{\Gamma_{7}}^{\beta \dagger}(i \Omega_{m},\bm{q}) \Big\rangle  \nonumber \\
&& = \Bigg\{ \Big(\Omega_{m}^{2} + \Omega_{\bm{q}}^{2}\Big) \delta_{\alpha\beta}
- \frac{\lambda^{2}}{\Omega_{\bm{q}}} \Bigg(\frac{\delta_{\alpha\beta}}{V}
\Big(\frac{\delta_{\beta\beta'}}{V} + \Pi^{\beta\beta'} \Big)^{-1}
\Pi^{\beta'\beta} \nonumber \\
&& -  \Pi^{\alpha\beta}
\Bigg( \frac{\delta_{\beta\beta'}}{V}
+ \frac{\delta_{\beta\gamma}}{V}
\Big(\frac{\delta_{\gamma\gamma'}}{V} + \Pi^{\gamma'\beta'} \Big)^{-1}
\Pi^{\beta'\beta} \Bigg)^{-1}
\Pi^{\beta'\beta}\Bigg) \Bigg\}^{-1} .
\end{eqnarray}
Here, $V$ is the bare interband Coulomb interaction, $\lambda$ is the electron--phonon coupling, and $\Omega_{\mathbf{q}}$ is the bare phonon frequency.
The indices $\alpha$ and $\beta$ label the internal components of the coupled $\Gamma_4$ and $\Gamma_7$ sectors, and all matrix inversions act within this internal space.

The nested inverse structure of Eq.~(\ref{eq:phonon}) has a clear physical interpretation. The $\Gamma_7$ phonon self-energy is not determined by a single polarization bubble, but is instead dressed by repeated mode-conversion processes between the symmetry-incompatible $\Gamma_4$ and $\Gamma_7$ channels mediated by the same interband fluctuations. This feedback strongly enhances phonon softening even when the bare electron–phonon coupling is modest. The CDW transition is therefore determined by the instability condition that the lowest eigenvalue of the inverse propagator in Eq.~(\ref{eq:phonon}) vanishes at $q = Q_{CDW}$, thereby determining the phase boundary shown in Fig.~\ref{fig:main}.

\begin{figure*}
\includegraphics[width=\textwidth]{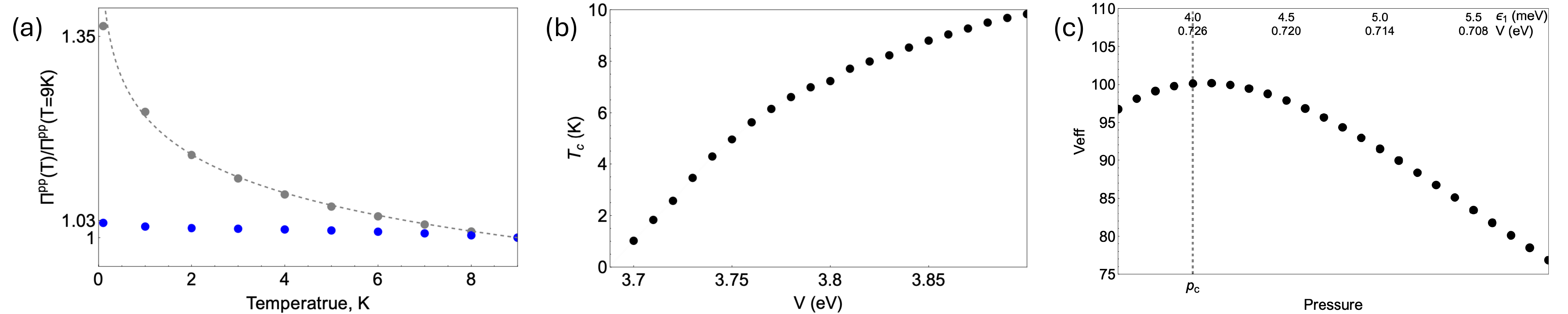}
\captionsetup{justification=raggedright}
\caption{
(a) Temperature dependence of the pair susceptibility $\Pi^{pp}(T)$.
The conventional intra-orbital $d$--$d$ pairing channel obtained from numerical calculations (gray dots) exhibits a logarithmic divergence (gray dashed line, $\sim \log(\Lambda/T)$), whereas the inter-orbital $p$--$d$ pairing susceptibility (blue dots) remains nearly constant down to the lowest accessible temperatures.
(b) Superconducting transition temperature $T_c$ as a function of the bare Coulomb interaction $V$ near the quantum critical point (QCP).
Below a critical value of $V$, superconductivity disappears.
This behavior contrasts with the expectation from conventional BCS theory, $T_c \sim T_F \exp[-1/(N_0 V)]$, reflecting the absence of a logarithmic divergence in the pair susceptibility.
(c) Effective interaction near the Fermi energy at $T = 10\,\mathrm{K}$ as a function of pressure.
The interaction peaks at the quantum critical point $p_c$ (dashed line).
The upper labels indicate the corresponding values of the $\Gamma$-band energy $\varepsilon_1$, which tracks the pressure-induced band shift, and the bare Coulomb interaction $V$, which decreases with increasing pressure.
The enhancement near $p_c$ gives rise to the dome-shaped superconducting phase diagram shown in Fig.~\ref{fig:main}.
}
\label{fig:pipp}
\end{figure*}

As the chiral CDW spectrum softens toward the quantum critical point, the effective interaction is strongly enhanced and drives superconductivity.
Because the low-energy states relevant to the CDW instability are orbitally distinct and connected by $Q_{CDW}$, the leading superconducting instability is necessarily inter-orbital.
This kinematic constraint enforces a finite-momentum pairing instability and disfavors a reduction to conventional zero-momentum intraband pairing.
This distinction is reflected in the contrasting temperature dependence of the inter-orbital and intra-orbital pair susceptibilities, highlighting the absence of a Cooper logarithm in the dominant pairing channel and its implications for the superconducting dome.

Figure~\ref{fig:pipp} reveals a qualitative distinction between the dominant inter-orbital $p$--$d$ pairing channel and conventional intra-orbital pairing.
While the intra-orbital $d$--$d$ pair susceptibility exhibits the familiar low-temperature enhancement associated with the Cooper logarithm, the inter-orbital $\Gamma$--$L$ susceptibility remains nearly temperature independent down to the lowest accessible temperatures.
This striking contrast reflects a fundamental difference in the available pairing phase space.
The inter-orbital pair susceptibility is defined as
\begin{equation}
\begin{aligned}
\Pi_{\alpha\alpha'}^{pp}(i \Omega_{m},\bm{q})
&= \frac{1}{\beta} \sum_{i \omega_{n}}
\int \frac{d^{3} \bm{k}}{(2\pi)^{3}} \\
&\quad \times
G^{\alpha\beta}_{\Gamma}(i \omega_{n}+i\Omega_{m},\bm{k}+\bm{q})
G^{\beta\alpha'}_{L}(-i \omega_{n},-\bm{k}) .
\end{aligned}
\label{eq:pairsus}
\end{equation}
In conventional BCS theory, the Cooper logarithm arises from an extended manifold of momenta that simultaneously satisfy the on-shell condition for both propagators near a common Fermi surface.
This joint on-shell condition generates a logarithmic enhancement of the pair susceptibility at low temperature.
In contrast, in pressurized TiSe$_2$ the relevant $\Gamma$ and $L$ pockets are small, orbitally distinct, and displaced in momentum space, communicating primarily through the finite ordering vector $Q_{\mathrm{CDW}}$.
As a result, the set of momenta that satisfy the simultaneous on-shell condition for both Green's functions is strongly restricted.
The associated phase space is therefore insufficient to generate a Cooper logarithm.
Superconductivity in this regime is consequently controlled by the strength of the effective interaction rather than by a density-of-states-driven instability.

The superconducting transition is determined by the linearized instability condition near the Fermi energy,
\begin{eqnarray}\label{eq:cooper}
    1 = \Pi_{\alpha\alpha'}^{pp}(0,\bm{q})\, V_{\mathrm{eff}}(0,\bm{q}) .
\end{eqnarray}
Here $V_{\mathrm{eff}}$ denotes the effective pairing interaction mediated by the renormalized phonon mode and the chiral CDW collective fluctuations, obtained from the same RPA treatment used to describe the CDW sector.
The explicit expression for $V_{\mathrm{eff}}$ and the definitions of the intermediate propagators are provided in Supplementary Information Section III.\cite{SM}

A direct consequence of the nondivergent behavior of $\Pi^{pp}$ is the emergence of a threshold interaction strength for superconductivity. As shown in Fig.~\ref{fig:pipp}b, the superconducting transition temperature $T_c$ vanishes below a critical value $V_c$, in contrast to the weak-coupling BCS scenario where any attractive interaction leads to pairing through a logarithmic divergence. Instead, the superconducting instability is controlled by whether the effective interaction exceeds a finite threshold set by the nondivergent pair susceptibility. In this regime, a dome-shaped superconducting phase can naturally arise from the enhancement of $V_{\mathrm{eff}}$ near the CDW quantum critical point, even when the temperature dependence of $\Pi^{pp}(T)$ remains weak.

Figure~\ref{fig:pipp}c reveals the microscopic origin of the superconducting dome through the nonmonotonic evolution of the effective interaction $V_{\mathrm{eff}}$. The effective interaction near the Fermi energy is maximized at the chiral CDW quantum critical point and decreases on either side. On the high-pressure side, the weakening of chiral CDW fluctuations suppresses $V_{\mathrm{eff}}$, while on the low-pressure side the onset of long-range CDW order competes with superconductivity by reducing the low-energy fluctuations available for pairing.
Both effects are naturally incorporated into $V_{\mathrm{eff}}$ through the collective-mode structure captured by the RPA resummation. Their combined influence gives rise to the dome-shaped superconducting phase near the endpoint of the chiral CDW phase shown in Fig.~\ref{fig:main}.

We next analyze the symmetry of the resulting inter-orbital Cooper pairs using a group-theoretical approach tailored to orbital-mixed pairing. Similar symmetry-based analyses have been employed in other multi-orbital superconductors, such as strontium ruthenate,\cite{Kaba2019} where the orbital character plays a central role in constraining the allowed gap structures.
Applying this framework to TiSe$_2$, we focus on inter-orbital $p$–$d$ pairing within the $D_{3d}$ point group. Owing to the odd parity of the $p$ orbital, the symmetry classification of the $p$–$d$ Cooper pair is nontrivial and imposes additional constraints on the allowed pairing channels.
Combining the group-theoretical constraints with a tight-binding description of the electronic structure, our analysis indicates that an orbital-selective $s$-wave pairing state is a natural candidate for the superconducting order, while nodal pairing states appear less favorable within the parameter regime considered here.
This conclusion is consistent with experimental observations of nodeless superconductivity in TiSe$_2$.\cite{Li2007,Zaberchik2010}
Details of the group-theoretical analysis and the tight-binding calculations are provided in Methods and Supplementary Information IV.\cite{SM}

\section{Discussion}

Building on the finite-center-of-mass-momentum inter-orbital pairing mechanism established above, we now examine its broader physical implications and experimentally testable consequences.
The absence of a logarithmically divergent Cooper susceptibility implies that superconductivity in pressurized TiSe$_2$ is interaction-driven and kinematically constrained, placing it outside the conventional intraband BCS paradigm.

This mechanism leads to concrete and falsifiable experimental predictions.
Because Cooper pairs carry the ordering vector $Q_{\mathrm{CDW}}$, superconducting correlations inherit the momentum structure of the chiral charge-density-wave sector.
Phase-sensitive Josephson experiments in junctions that approximately conserve in-plane momentum provide a direct probe.
In such geometries, the critical current is expected to acquire an orientation dependence reflecting projection onto the finite-$Q$ channel, a feature absent in zero-momentum superconductors.\cite{Chen2023nature,Hamidian2016,Du2020,Zhao2023,Liu2023,Chen2021,Aishwarya2023,Gu2023,Liu2021,Wu2023,Setty2023,Agterberg2020,Fradkin2015}

A second prediction concerns disorder.
Because pairing relies on coherent coupling between $\Gamma$ and $L$ pockets separated by $Q_{\mathrm{CDW}}$, scattering processes that relax momentum on this scale should suppress superconductivity more strongly than in intraband BCS systems.
Controlled comparisons of pressure-tuned samples with varying impurity concentrations would therefore test the momentum-selective nature of pairing.
A third signature follows from electronic-structure evolution.
Superconductivity should track the pressure-induced reconstruction of the $\Gamma$-centered hole pocket near the Lifshitz transition that coincides with the superconducting dome.\cite{Hinlopen2024}
Simultaneous measurements of Fermi-surface topology and $T_c$ under pressure would directly probe the kinematic origin of pairing.

Together these criteria distinguish interaction-driven finite-$Q$ inter-orbital superconductivity from conventional zero-momentum intraband pairing in TiSe$_2$.
More broadly, the underlying ingredients are not unique to TiSe$_2$.
Whenever small, orbitally distinct Fermi pockets are connected by a charge-density-wave wave vector and reconstructed across a symmetry-breaking transition, critical fluctuations can naturally promote finite-momentum inter-orbital pairing.
TiSe$_2$ thus provides a concrete realization of a broader route to superconductivity beyond conventional Fermi-surface instabilities.

We next examine why conventional intraband pairing or dominant pair-hopping mechanisms are unlikely to govern superconductivity in the present setting.
Because the relevant Fermi pockets remain small near the quantum critical regime, an intraband instability would require either a substantially enhanced density of states or an unrealistically strong effective attraction to account for the observed $T_c \sim 2$--$4$~K.
Our calculations indicate that reproducing $T_c \approx 3$~K via intraband pairing in the $\Gamma$-centered hole band would require an interaction strength on the order of 5~eV, which is implausibly large for TiSe$_2$, consistent with first-principles estimates placing a purely phonon-mediated $T_c$ below 1~K.\cite{Calandra2011}
Although intraband channels are not excluded within our formalism, their effectiveness is limited by the momentum structure of the collective modes.
Phonons at $\mathbf{q}=0$ do not soften near the chiral CDW quantum critical point, whereas the effective interaction is selectively amplified by soft fluctuations at $Q_{\mathrm{CDW}}$.
This momentum-selective critical enhancement naturally renders intraband coupling subleading to the inter-orbital $\Gamma$--$L$ channel.

Although the presence of multiple small Fermi pockets may invite comparison to $s^{\pm}$ scenarios proposed for iron-based superconductors,\cite{Sprau2017} the microscopic situation in TiSe$_2$ is qualitatively different.
The pressure-induced $\Gamma$ pocket is predominantly Se $p$-like, whereas the $L$ pockets are Ti $d$-derived, leading to a pronounced orbital mismatch.
Because coherent pair-hopping amplitudes scale with orbital overlap, this $p$–$d$ disparity naturally suppresses inter-pocket pair transfer.
As a result, mechanisms based on near-perfect nesting\cite{Ganesh2014} or dominant pair hopping become strongly model dependent in this material.
These considerations instead highlight the inter-orbital $\Gamma$--$L$ channel, amplified by chiral CDW critical fluctuations, as the leading superconducting instability in TiSe$_2$.
In this broader context, renormalization-group studies of single-layer TiSe$_2$ have proposed electronically driven unconventional superconducting states arising from inter-pocket Coulomb interactions~\cite{Ganesh2014}; these scenarios, formulated for a two-dimensional system with zero-momentum pairing, differ from the finite-momentum inter-orbital mechanism discussed here in the bulk material near a chiral CDW quantum critical point.

\begin{figure}
\includegraphics[width=\columnwidth]{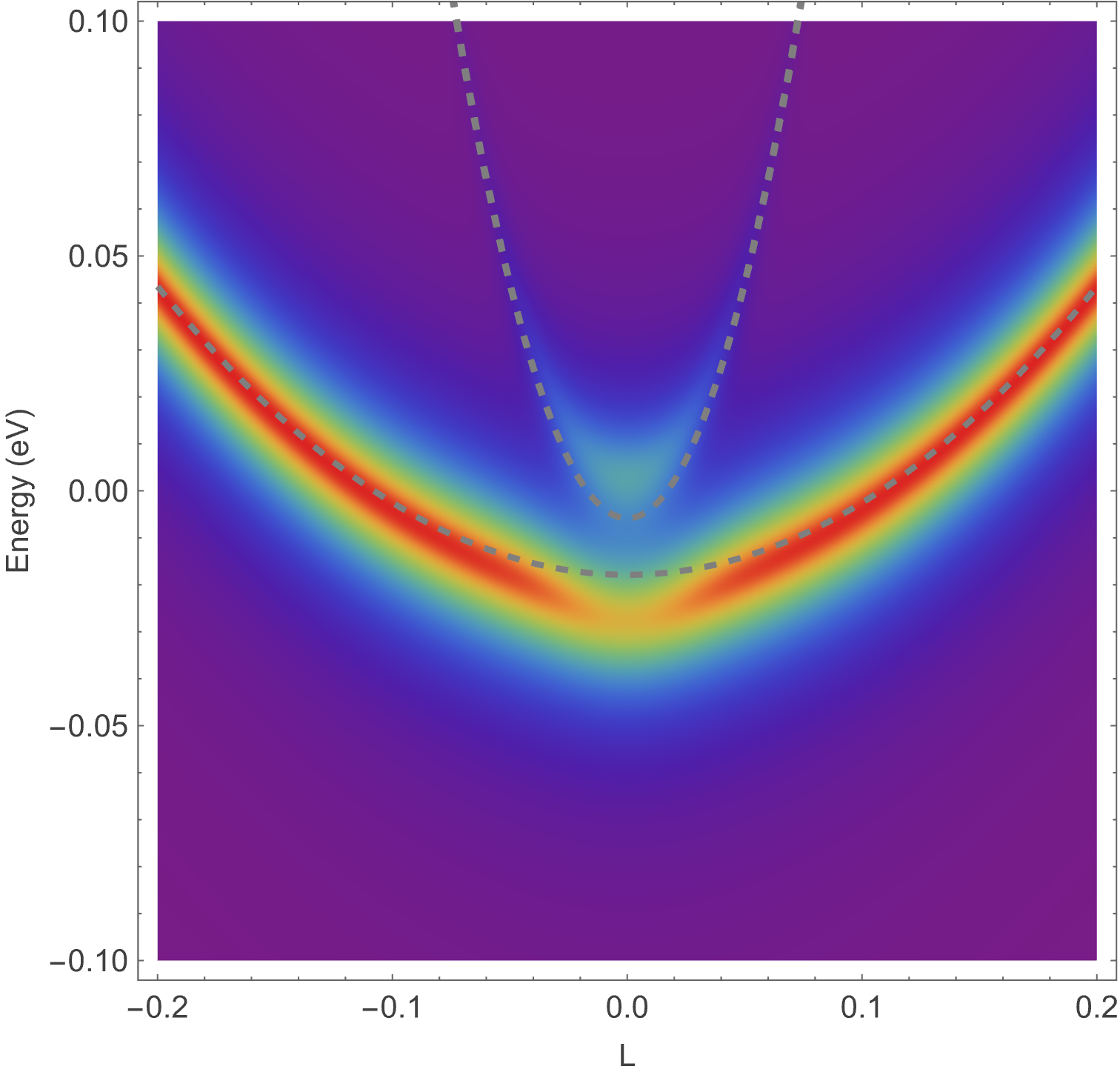}
\captionsetup{justification=raggedright}
\caption{
Gapless quasiparticle dispersion in the superconducting state near the $L$ point.
The calculated spectral function reflects inter-orbital finite-momentum pairing between the $\Gamma$- and $L$-centered states.
The lower (upper) dashed line denotes the normal-state conduction band near $L$ (the $\Gamma$-derived hole band folded to $L$ by $Q_{\mathrm{CDW}}$).
No clear full gap is resolved at the Fermi level.
Instead, superconducting signatures are more pronounced in the reconstructed $\Gamma$-derived band, while the $L$-centered conduction band exhibits a suppression of spectral weight without a well-defined gap.
A complementary view near the $\Gamma$ point is shown in Supplementary Fig.~S1.
}
\label{fig:gapless}
\end{figure}

This exclusion of conventional intraband pairing mechanisms has direct implications for the nature of the superconducting state.
In the inter-orbital $\Gamma$–$L$ pairing scenario, superconductivity in pressurized TiSe$_2$ is not expected to manifest as a conventional gap opening on a single Fermi surface.
As illustrated schematically in Fig.~\ref{fig:gapless}, Cooper pairs form between conduction- and valence-band states rather than within a single band crossing the Fermi level, and the dominant pairing channel therefore lacks a logarithmically divergent Cooper susceptibility, consistent with the nearly temperature-independent $p$–$d$ pair susceptibility shown in Fig.~\ref{fig:pipp}.

Such a pairing mechanism implies a superconducting state that is comparatively fragile, as it does not originate from a sharply defined Fermi-surface instability.
This perspective aligns with the narrow superconducting dome and the pronounced sensitivity of TiSe$_2$ to pressure and carrier doping.
Perturbations that disrupt coherent coupling between the $\Gamma$ and $L$ sectors, or relax momentum selectivity at $Q_{\mathrm{CDW}}$, are therefore expected to exert an amplified influence on pairing.
This character is also reflected in the electronic spectral function: the redistribution and partial suppression of spectral weight shown in Fig.~\ref{fig:gapless} are compatible with reduced quasiparticle coherence rather than a conventional Fermi-surface gap opening.
A definitive distinction between incoherent spectral suppression and a symmetry-resolved superconducting gap, however, will require further momentum-resolved spectroscopic studies.

The considerations discussed above place superconductivity in TiSe$_2$ within a quantum critical framework, in which pairing interactions are enhanced by critical fluctuations of the chiral CDW order.
This perspective does not exclude the complementary scenario in which superconductivity nucleates at domain walls associated with incommensurate CDWs,\cite{Joe2014,Li2016,Kogar2017prl} but instead indicates that quantum critical fluctuations provide the dominant pairing channel, with domain-wall physics potentially contributing at a more microscopic level.
Clarifying the relative roles of these mechanisms requires a well-established phase diagram.
Early reports indicated that the superconducting dome emerges at the critical doping $x_c$ in Cu intercalation, coincident with the termination of the commensurate CDW (CCDW),\cite{Joe2014,Kogar2017prl} whereas under pressure superconductivity was initially reported to appear within the CCDW phase,\cite{Kusmartseva2009} with $p_c$ marking its end point.
More recent high-pressure measurements, however, suggest that the dome can also develop near $p_c$,\cite{Hinlopen2024} highlighting that the pressure phase diagram remains unsettled.

Although our analysis emphasizes the quantum critical scenario, the distinct electronic structures and carrier characteristics observed at $x_c$ and $p_c$ indicate that superconductivity may not originate from an identical mechanism in the doping- and pressure-tuned regimes.
Resolving this issue requires experiments that directly probe the superconducting state, including detailed characterization of the gap structure, magnetic-field response, and impurity effects.
In particular, the finite-momentum inter-orbital $\Gamma$–$L$ channel identified here motivates phase-sensitive and momentum-resolved measurements targeting pairing correlations at the CDW wave vector $Q_{\mathrm{CDW}}$.\cite{Chen2023nature,Hamidian2016,Du2020,Zhao2023,Liu2023,Chen2021,Aishwarya2023,Gu2023,Liu2021,Wu2023,Setty2023,Agterberg2020,Fradkin2015}
More broadly, our results outline a general route to superconductivity in chiral CDW materials with reconstructed multi-orbital electronic structures.
In such systems, critical CDW fluctuations can naturally promote finite-momentum inter-orbital pairing, providing a unified framework that extends beyond TiSe$_2$.

\section{Method}

\subsection{Effective Hamiltonian for the electron dynamics}
We model the low-energy electronic structure of TiSe$_2$ near the $\Gamma$ point and the three symmetry-related $L$ points using a symmetry-constrained $\mathbf{k}\!\cdot\!\mathbf{p}$ Hamiltonian for the $D_{3d}$ point group following Ref.~\cite{Huang2021}.
The $\Gamma$ sector consists of three Se $p$-derived states, while the $L$ sector is represented by a Ti $t_{2g}$-derived conduction band capturing the relevant electron pocket.
All matrix conventions and parameter values are summarized in Supplementary Section~I.

Pressure is incorporated phenomenologically at the level of the low-energy electronic structure.
Specifically, we implement pressure as an upward shift of the $\Gamma$-centered hole sector relative to the chemical potential, consistent with the experimentally observed Lifshitz transition.
Throughout this work, the ordering wave vector is fixed to the experimentally observed $Q_{\mathrm{CDW}}$.

\subsection{Effective field theory for chiral CDW}
The chiral CDW sector is formulated in terms of interband particle--hole fluctuations between the $\Gamma$ and $L$ pockets.
The corresponding polarization bubble $\Pi^{\alpha\beta}(i\Omega_m,\mathbf{q})$ is evaluated from the $\Gamma$ and $L$ Green’s functions of the effective electronic model.
Both the momentum integration and the Matsubara frequency summation are performed numerically, with cutoffs chosen to ensure convergence of the polarization function.
The internal indices $\alpha$ and $\beta$ label the components of the coupled $\Gamma_4$ and $\Gamma_7$ sectors introduced in the main text.
Details of the discretization scheme and convergence checks are provided in Supplementary Section~II.

The renormalized collective-mode propagator is obtained within a random-phase-approximation (RPA) resummation that combines the bare interband Coulomb interaction $V$, the electron--phonon coupling $\lambda$, and the bare phonon dispersion $\Omega_{\mathbf{q}}$ with the polarization matrix $\Pi$.
The CDW transition temperature $T_{\mathrm{CDW}}$ is determined by locating the temperature at which the lowest eigenvalue of the inverse renormalized propagator at $\mathbf{q}=Q_{\mathrm{CDW}}$ vanishes.

Pressure enters the RPA analysis through the screened interband Coulomb interaction.
In practice, we take the bare interaction parameter $V$ to decrease linearly with pressure, reflecting enhanced electronic screening under compression.\cite{Bok2021}
No specific functional form beyond this linear dependence is assumed, and the qualitative features discussed in this work are insensitive to the precise slope of $V(p)$.
Importantly, although $V$ decreases monotonically with pressure, the emergence of quantum criticality and the enhancement of superconductivity arise from the nontrivial renormalization of the coupled collective modes within the RPA framework.

\subsection{Effective field theory for superconductivity}
Superconducting instabilities are analyzed in the inter-orbital $\Gamma$--$L$ channel.
The pair susceptibility $\Pi^{pp}$ is defined as in Eq.~(\ref{eq:pairsus}) and evaluated using the same low-energy Green’s functions as in the CDW sector.
Both the Matsubara frequency summation and the momentum integration are performed numerically, using identical cutoffs and discretization schemes to ensure consistency between the particle--hole and particle--particle channels.
The resulting inter-orbital pair susceptibility exhibits only a weak temperature dependence, reflecting the absence of a Cooper logarithm.

The effective pairing interaction $V_{\mathrm{eff}}$ is constructed from the exchange of RPA-renormalized collective modes.
This includes contributions from both the renormalized phonon propagator and the electronic chiral CDW collective fluctuations in the coupled $\Gamma_4$ and $\Gamma_7$ sectors.
Explicit expressions for $V_{\mathrm{eff}}$ and all intermediate propagators are provided in Supplementary Section~III.

The superconducting transition temperature $T_c$ is determined from the leading eigenvalue of the linearized pairing kernel, which reduces to the scalar instability condition in Eq.~(\ref{eq:cooper}) for the dominant inter-orbital channel.
In Fig.~\ref{fig:pipp}b, the bare interaction $V$ is varied at fixed pressure near the quantum critical point to illustrate the threshold nature of superconductivity.
In contrast, the pressure dependence of $T_c$ shown in Fig.~\ref{fig:main} follows from the combined evolution of the electronic structure, the collective-mode spectrum, and the linearly reduced interaction parameter $V(p)$.

\subsection{Representation analysis for pairing symmetry}
We classify inter orbital $p$ to $d$ pairing bilinears in the $D_{3d}$ point group using a basis of three Se $p$ orbitals and three Ti $t_{2g}$ orbitals.
The pairing matrix is decomposed into irreducible representations of $D_{3d}$ subject to Fermi statistics and the orbital parity structure of $p$ and $d$ states.
We then project the allowed symmetry channels onto the low energy $\Gamma$ and $L$ band states of the effective Hamiltonian to identify the leading candidate gap structure.
All representation matrices and projection formulas are provided in Supplementary Section IV.

\subsection{Calculation of superconducting spectral function}
To compute the quasiparticle spectrum in the superconducting state, we evaluate the single-particle Green’s function associated with inter-orbital finite-momentum pairing between the hole bands near the $\Gamma$ point and the conduction bands near the $L$ points. The $\Gamma$-point hole sector is described by a three-band Hamiltonian $H^\Gamma(\mathbf{k})$, whose explicit form and parameters are given in the Supplementary Information.
For Fig.~\ref{fig:gapless}, all three hole bands at $\Gamma$ are retained, while only a single representative conduction band at one $L$ point is included. The remaining two $L$ points are related by symmetry and yield identical spectral features.
In the Nambu basis
\begin{equation}
\Psi_{\mathbf{k}} =
\begin{pmatrix}
\psi_{\Gamma}(\mathbf{k}) \\
c^{\dagger}_{L,\mathbf{Q}_{\mathrm{CDW}}-\mathbf{k}}
\end{pmatrix},
\qquad
\psi_{\Gamma}(\mathbf{k})=
\begin{pmatrix}
c_{\Gamma 1,\mathbf{k}}\\
c_{\Gamma 2,\mathbf{k}}\\
c_{\Gamma 3,\mathbf{k}}
\end{pmatrix},
\end{equation}
the inverse Green’s function is written in block form as
\begin{align}
G^{-1}(\mathbf{k}, i\omega_n)
&= i\omega_n \mathbb{I}_4 - H(\mathbf{k}) \notag \\
&=
\begin{pmatrix}
i\omega_n\,\mathbb{I}_3 - H^\Gamma(\mathbf{k})
& \Delta_{\Gamma L} \\
\Delta_{\Gamma L}^{\dagger}
& i\omega_n + \xi_{L}(\mathbf{Q}_{\mathrm{CDW}}-\mathbf{k})
\end{pmatrix}.
\end{align}

For Fig.~\ref{fig:gapless}, the magnitude of the inter-orbital pairing amplitude is set to $|\boldsymbol{\Delta}_{\Gamma L}| = 10$~meV. This choice is made to clearly resolve the qualitative effects of inter-orbital pairing on the quasiparticle dispersion and does not imply the physical superconducting gap scale in TiSe$_2$. The retarded Green’s function is obtained by analytic continuation $i\omega_n \rightarrow \omega + i\eta$, where $\eta$ is a small phenomenological broadening parameter. The momentum-resolved spectral function is evaluated as
\begin{equation}
A(\mathbf{k},\omega)
=
-\frac{1}{\pi}
\,\mathrm{Im}\,
\mathrm{Tr}\,
G^{R}(\mathbf{k},\omega).
\end{equation}
The dispersion shown in Fig.~\ref{fig:gapless} is obtained by evaluating $A(\mathbf{k},\omega)$ along a momentum cut passing through the selected $L$ point.

\section{Acknowledgements}
\begin{acknowledgments}
K.-S. Kim was supported by the Ministry of Education, Science, and Technology (RS-2024-00337134) of the National Research Foundation of Korea (NRF) and by TJ Park Science Fellowship of the POSCO TJ Park Foundation. J. M. B. acknowledges support from the National Research Foundation of Korea (NRF) (No. RS-2022-NR072365).
\end{acknowledgments}

\clearpage
\onecolumngrid

\setcounter{equation}{0}
\renewcommand{\theequation}{S\arabic{equation}}

\setcounter{table}{0}
\renewcommand{\thetable}{S\arabic{table}}

\setcounter{figure}{0}
\renewcommand{\thefigure}{S\arabic{figure}}

\makeatletter
\setcounter{NAT@ctr}{0}
\makeatother

\begin{center}
{\Large\bfseries Supplementary Information for}\\[0.6em]
{\large\itshape Finite-momentum inter-orbital superconductivity driven by chiral charge-density-wave quantum criticality beyond the BCS regime}\\[1.0em]
Jin Mo Bok$^{1}$, B. J. Kim$^{1}$, Ki-Seok Kim$^{1}$\\[0.4em]
$^{1}$Department of Physics, Pohang University of Science and Technology, Pohang 37673, South Korea\\[0.2em]
{\ttfamily jinmobok@postech.ac.kr, tkfkd@postech.ac.kr}
\end{center}

\vspace{0.8em}

\section{Effective Hamiltonian for the electron dynamics}

In this section we specify the complete low-energy model used throughout the work, including the basis choice, the ordering of internal indices, and the explicit $3\times3$ matrix forms at $\Gamma$ and at each $L$ point \cite{SM_Huang2021}.
We list the parameter values used in the calculations and specify how pressure is incorporated in the effective theory.
In the calculations, pressure is not treated as an independent external variable but is implemented phenomenologically by tuning the $\Gamma$-band energy $\varepsilon_1$, corresponding to an upward shift of the $\Gamma$-centered hole bands, while all other parameters are kept fixed unless otherwise stated.
Strictly speaking, such a band shift should be accompanied by a chemical-potential adjustment to preserve charge neutrality; here we neglect this effect and treat the band shift as an effective parametrization of pressure.
This approximation is justified by the fact that superconductivity in the present system is governed primarily by interaction effects mediated by enhanced CDW fluctuations rather than by fine details of the electronic density of states, and we therefore expect that a small charge imbalance does not qualitatively affect the conclusions of this work.
We also document the gapped $\Gamma$–$L$ spectrum choice and the resulting Fermi-surface evolution that serves as input for the RPA and pairing calculations.

In the following we retain only the bands around $\Gamma$ and the three equivalent $L$ points which participate in the charge density wave and superconducting instabilities and we omit the bands near $M$ that do not enter our low-energy analysis.
The imaginary time action for non-interacting electrons $S_{el}$ together with the corresponding $3\times3$ matrix Hamiltonians $H^{\Gamma}$ and $H^L$ at the $\Gamma$ and $L$ points are written as

\begin{eqnarray}
    &&S_{el}(\psi_{\Gamma_{\alpha}}(i\omega_n,\mathbf{k}),\psi_{L_{\alpha}}(i\omega_n,\mathbf{k})) \nn
    &&=\frac{1}{\beta}\sum_{i\omega_n}\int\frac{d^3\mathbf{k}}{(2\pi)^3}\{\psi^{\dagger}_{\Gamma_{\alpha}}(i\omega_n,\mathbf{k})(-i\omega_n\delta_{\alpha\beta}+H^{\Gamma}_{\alpha\beta}(\mathbf{k}))\psi_{\Gamma_{\beta}}(i\omega_n,\mathbf{k}) \nn
    &&+\psi^{\dagger}_{L_{\alpha}}(i\omega_n,\mathbf{k})(-i\omega_n\delta_{\alpha\beta}+H^{L}_{\alpha\beta}(\mathbf{k}))\psi_{L_{\beta}}(i\omega_n,\mathbf{k})\}
    \label{eq:action_el}
\end{eqnarray}
\begin{eqnarray}
     &&H^\Gamma(\mathbf{k}) \nonumber \\
    && =
 \begin{pmatrix}
     ak^2_\parallel +a'k^2_z +\varepsilon_1 + c k_y k_z + c'(k^2_x-k^2_y) & -c k_x k_z -2 c' k_x k_y & i\sqrt{2}d k_y \\
     -c k_x k_z -2 c' k_x k_y & ak^2_\parallel +a'k^2_z +\varepsilon_1 - c k_y k_z - c'(k^2_x-k^2_y) & i\sqrt{2}d k_x \\
     -i\sqrt{2}d k_y  & -i\sqrt{2}d k_x & b k^2_\parallel + b' k^2_z +\varepsilon_2
 \end{pmatrix}, \nonumber \\
\end{eqnarray}

\begin{equation}
 H^{L}(\mathbf{k})=
 \begin{pmatrix}
    \xi^{L}_1(\mathbf{k}) & 0 & 0 \\
    0 &  \xi^{L}_2(\mathbf{k}) & 0 \\
    0 & 0 & \xi^{L}_3(\mathbf{k})
 \end{pmatrix},
\end{equation}

\begin{equation}
    \xi^{L}_2(\mathbf{k})=A^{L}k^2_x+B^{L}k^2_y+C^{L}k^2_z+D k_y k_z + \varepsilon^{L}_3.
\end{equation}

Here $k^2_\parallel=k^2_x+k^2_y$ and $\xi_1(\mathbf{k})$ and $\xi_3(\mathbf{k})$ are obtained by rotating $\xi_2(\mathbf{k})$ by $\pm 120^\circ$ around the $k_z$ axis.
At the $\Gamma$ point the two bands close to the Fermi level form an $E_\mathrm{g}$ doublet and the deeper band has $A_\mathrm{2u}$ symmetry while the bands at the three symmetry related $L_{i=1,2,3}$ points carry $A_\mathrm{g}$ symmetry and cross the Fermi level slightly.
The parameters in $H^{\Gamma}$ and $H^L$ are obtained by fitting to the tight-binding model derived from density functional theory calculations by Monney $\textit{et al}$.\cite{SM_Monney2011dft} with emphasis on reproducing the dispersions near the high symmetry points $\Gamma$ and $L$ as shown in Fig.2 of main text. The energies $\varepsilon_1=-0.008$ eV, $\varepsilon_2=-0.147$ eV and $\varepsilon^L_3=-0.004$ eV are taken directly from the tight-binding model and the remaining coefficients are summarized in Table. \ref{table:dispersion}.
The small values of $c$ and $c'$ in $H^{\Gamma}$ reflect the near degeneracy of the $E_\mathrm{g}$ doublet at $\Gamma$ and the splitting at about $-0.15$ eV which lies well below the Fermi level in the tight-binding bands is neglected.
With this choice of parameters the effective model exhibits a small indirect gap between the $\Gamma$ centered hole band and the $L$ centered electron band providing a consistent low-energy starting point for the analyses presented in this Supplementary Information and in the main text.
To motivate this choice we note that ARPES measurements on pristine TiSe$_2$ resolve a small but finite indirect gap between the top of the Se $p$ valence band at $\Gamma$ and the bottom of the Ti $t_{2g}$ conduction band at $L$ in the normal state so the gapped model provides a more realistic starting point for our low-energy analysis.\cite{SM_Rasch2008,SM_May2011,SM_Chen2015}
In particular, this gapped $\Gamma$–$L$ spectrum determines the presence or absence of Fermi pockets at $\Gamma$ and $L$ as pressure is tuned, and the resulting Fermi-surface evolution constitutes the electronic input for both the RPA analysis of CDW fluctuations and the subsequent pairing calculations.

\begin{table}
\centering
\begin{ruledtabular}
    \begin{tabular}{cccc}
       & Bands  &  Parameters  \\ \hline \\
       & $H^{\Gamma}$  &  $a=-20.0\,\mathrm{eV}\mathrm{\AA}^{2}$, $a'=0.04\,\mathrm{eV}\mathrm{\AA}^{2}$,\\ & & $b=-3\,\mathrm{eV}\mathrm{\AA}^{2}$, $b'=-19.2\,\mathrm{eV}\mathrm{\AA}^{2}$,\\ & & $d=-0.007\,\mathrm{eV}\mathrm{\AA}$ \\  \\
       & $H^L$  & $A^L=6.3\,\mathrm{eV}\mathrm{\AA}^{2}$, $B^L=1.5\,\mathrm{eV}\mathrm{\AA}^{2}$, \\ & & $C^L=0.24\,\mathrm{eV}\mathrm{\AA}^{2}$, $D^L=0.0\,\mathrm{eV}\mathrm{\AA}^{2}$ \\ \\
	\end{tabular}
\end{ruledtabular}
\captionsetup{justification=raggedright}
\caption{The fitting parameters for $H^{\Gamma}$ and $H^L$.}
\label{table:dispersion}
\end{table}

\section{Effective field theory for chiral CDW}

In this section we provide a self-contained derivation of the coupled $\Gamma_4$ and $\Gamma_7$ fluctuation theory at $q = Q_{CDW}$ with all fields and normalization conventions defined explicitly.
We present the matrix structure of the RPA inversion and the precise form of the polarization bubble $\Pi$ that mixes the electronic $\Gamma_4$ sector with the $\Gamma_7$ phonon sector.
This material supplies the exact definitions behind Eq.2 in the main text and is intended to enable full reproduction of the renormalized collective mode propagator.

We begin by defining the interband CDW order parameter in the $\Gamma\leftrightarrow L_{i=1,2,3}$ channel.
\begin{eqnarray}
&&\phi^\dagger(i\Omega_m,\mathbf{q})\nonumber\\&&=\frac{1}{\beta}\sum_{i\omega_n}\int\frac{d^3\mathbf{k}}{(2\pi)^3}\psi^\dagger_{\Gamma}(i\omega_n+i\Omega_m,\mathbf{k}+\mathbf{q})\psi_{L}(i\omega_n,\mathbf{k})\nonumber\\
\end{eqnarray}
Group theoretical analysis shows that the three L points form an orbit of the $D_{\text{3d}}$ point group and that $\phi_{i}(i \Omega_{m},\bm{q})$ decomposes into eight three dimensional irreducible representations.
In particular the $\Gamma\leftrightarrow L_{i=1,2,3}$ CDW order parameter can be written as
\begin{eqnarray}
&& \phi_{i}(i \Omega_{m},\bm{q}) \nonumber \\ &&= \phi_{\Gamma_{7}}^{\alpha}(i \Omega_{m},\bm{q}) \oplus \phi_{\Gamma_{4}}^{\alpha}(i \Omega_{m},\bm{q}) \oplus \phi_{\Gamma_{1}}^{\alpha}(i \Omega_{m},\bm{q}) \oplus \phi_{\Gamma_{6}}^{\alpha}(i \Omega_{m},\bm{q}).\nonumber \\
\end{eqnarray}
The remaining irreps belong to the $\Gamma\leftrightarrow M_{i=1,2,3}$ channel.
Previous work showed that the $\Gamma\leftrightarrow L_{i=1,2,3}$ channel provides the leading contribution to the $2\times2\times2$ CDW order in TiSe$_2$ so we neglect the $\Gamma\leftrightarrow M_{i=1,2,3}$ channel in what follows.\cite{SM_Kim2024}

The $\Gamma_5$ irrep is associated with a $2\times2$ CDW in the $\Gamma\leftrightarrow M_{i=1,2,3}$ channel.
Although the product representation $\Gamma_7\otimes\Gamma_4$ yields $\Gamma_5$, this does not correspond to a scalar invariant at the $\Gamma$ point and therefore does not allow a direct linear coupling between the $\Gamma_4$ and $\Gamma_7$ modes.
Instead, the $\Gamma_5$ order is generated as a secondary order parameter, consistent with inelastic x-ray scattering observations.\cite{SM_Kim2024}
Scanning tunneling microscopy identifies the charge density modulation with the $\Gamma_4$ irrep and x ray diffraction identifies the soft phonon mode with the $\Gamma_7$ irrep.\cite{SM_Kim2024,SM_Ishioka2010}

The irreducible representation labels defined at the Brillouin zone center constrain only bilinear invariants at $\bm{q}=0$ and do not fix the structure of the theory at finite ordering wave vector.
At $\bm{q}=Q_{\mathrm{CDW}}$ the relevant symmetry is that of the little group of $Q_{\mathrm{CDW}}$, which allows additional bilinear terms absent at the zone center.
Therefore we allow a coupling between the charge ordering fluctuation associated with $\Gamma_4$ and the $\Gamma_7$ sector once the theory is formulated at $\bm{q}=Q_{\mathrm{CDW}}$.

Microscopically this mixing originates from interband $\Gamma$--$L$ particle hole fluctuations carrying momentum $Q_{\mathrm{CDW}}$ and is encoded in the polarization matrix $\Pi^{\alpha\beta}$.
This generates the effective mixing vertex
\begin{equation}\label{eq:comver_sm}
\phi_{\Gamma_4}^{\alpha\dagger}(\mathbf{q})\,\Pi^{\alpha\beta}(i\Omega_m,\mathbf{q})
\big[a_{\Gamma_7}^{\beta}(i\Omega_m,\mathbf{q})+\phi_{\Gamma_7}^{\beta}(i\Omega_m,\mathbf{q})\big].
\end{equation}
Orbital dependent vertex form factors are not treated explicitly and their net effect is absorbed into the effective coupling entering the subsequent RPA resummation.

Based on this symmetry analysis at $\bm{q}=Q_{\mathrm{CDW}}$ we retain only the $\Gamma_4$ and $\Gamma_7$ collective modes together with the $\Gamma_7$ phonon and construct the coupled effective action including electronic dynamics Coulomb interaction and electron phonon coupling.
The construction follows the coupled phonon CDW effective action introduced in Ref.\cite{SM_Kim2024}(Eq.(7)-(12) in supplementary information) and is summarized here for completeness.
Expanding this action to quadratic order in the collective fields $a_{\Gamma_7}$, $\phi_{\Gamma_7}$ and $\phi_{\Gamma_4}$ we obtain the effective action for the coupled phonon and CDW modes in matrix form.

\begin{multline}
S_{col}(a^{\alpha}_{\Gamma_7},\phi^{\alpha}_{\Gamma_7},\phi^{\alpha}_{\Gamma_4})
= \frac{1}{\beta}\sum_{i\Omega_m}\int\frac{d^3\mathbf{q}}{(2\pi)^3}\frac{1}{2}\,
\begin{pmatrix}
a^{\alpha\dagger}_{\Gamma_7} &
\phi^{\alpha\dagger}_{\Gamma_7} &
\phi^{\alpha\dagger}_{\Gamma_4}
\end{pmatrix}
\\
{\renewcommand{\arraystretch}{1.15}
\begin{pmatrix}
(\Omega_m^2+\Omega_{\mathbf{q}}^2)\delta_{\alpha\alpha'}
-\frac{\lambda^2}{\Omega_{\mathbf{q}}}\Pi^{\alpha\alpha'} &
-\frac{\lambda}{\sqrt{\Omega_{\mathbf{q}}}}\Pi^{\alpha\alpha'} &
-\frac{\lambda}{\sqrt{\Omega_{\mathbf{q}}}}\Pi^{\alpha\alpha'} \\
-\frac{\lambda}{\sqrt{\Omega_{\mathbf{q}}}}\Pi^{\alpha\alpha'} &
-\left(\frac{\delta_{\alpha\alpha'}}{V}+\Pi^{\alpha\alpha'}\right) &
-\Pi^{\alpha\alpha'} \\
-\frac{\lambda}{\sqrt{\Omega_{\mathbf{q}}}}\Pi^{\alpha\alpha'} &
-\Pi^{\alpha\alpha'} &
-\left(\frac{\delta_{\alpha\alpha'}}{V}+\Pi^{\alpha\alpha'}\right)
\end{pmatrix}}
\begin{pmatrix}
a^{\alpha'}_{\Gamma_7}\\
\phi^{\alpha'}_{\Gamma_7}\\
\phi^{\alpha'}_{\Gamma_4}
\end{pmatrix}.
\label{eq:collective}
\end{multline}
The off diagonal matrix elements proportional to $\Pi^{\alpha\alpha'}$ in Eq.~(\ref{eq:collective}) explicitly implement this fluctuation induced mixing between the $\Gamma_4$ and $\Gamma_7$ sectors.

Here $V$ denotes the bare interband Coulomb interaction, $\lambda$ is the electron-phonon coupling constant, and $\Omega_{\bm{q}}$ is the bare phonon frequency.
The matrix $\Pi^{\alpha\alpha'}(i\Omega_m,\bm{q})$ is the $\Gamma$-$L$ interband polarization bubble defined below, evaluated at $\bm{q}= Q_{\mathrm{CDW}}$.
The phonon field $a_{\Gamma_7}$ and the CDW collective modes $\phi_{\Gamma_7}$ and $\phi_{\Gamma_4}$ carry both frequency and momentum so their arguments are
\begin{eqnarray}
a^{\alpha^\dagger}_{\Gamma_7}=a^{\alpha^\dagger}_{\Gamma_7}(i\Omega_m,\mathbf{q}-\mathbf{q}_{\Gamma_7}), \\
\phi^{\alpha^\dagger}_{\Gamma_7}=\phi^{\alpha^\dagger}_{\Gamma_7}(i\Omega_m,\mathbf{q}-\mathbf{q}_{\Gamma_7}),\\
\phi^{\alpha^\dagger}_{\Gamma_4}=\phi^{\alpha^\dagger}_{\Gamma_4}(i\Omega_m,\mathbf{q}-\mathbf{q}_{\Gamma_4}).
\end{eqnarray}

The electronic polarization bubble that couples the $\Gamma$ and $L$ bands is given by
\begin{eqnarray}
    &&\Pi^{\alpha\alpha'}=\Pi^{\alpha\alpha'}(i\Omega_m,\mathbf{q})\nonumber\\ && = -\frac{1}{\beta}\sum_{i\omega_n,\mathbf{k}}G^{\alpha\beta}_{\Gamma}(i\omega_n+i\Omega_m,\mathbf{k+q})G^{\beta\alpha'}_{L}(i\omega_n,\mathbf{k})\nonumber\\
\end{eqnarray}

The Green functions are determined by the symmetry-constrained Hamiltonians $H^{\Gamma}$ and $H^L$ introduced in the previous subsection.
\begin{eqnarray}
    G^{\alpha\alpha'}_{\Gamma}(i\omega_n,\mathbf{k}) = (i\omega_n\delta_{\alpha\alpha'}-H^{\Gamma}_{\alpha\alpha'}(\mathbf{k}))^{-1} \\
    G^{\alpha\alpha'}_{L}(i\omega_n,\mathbf{k}) = (i\omega_n\delta_{\alpha\alpha'}-H^{L}_{\alpha\alpha'}(\mathbf{k}))^{-1}
\end{eqnarray}
Because the CDW instability is controlled by collective fluctuations at the ordering wave vector, we evaluate all polarization functions at $\bm{q} = Q_{\text{CDW}} = \mathbf{q}_{\Gamma_7} = \mathbf{q}_{\Gamma_4}$ and perform an RPA resummation of the Coulomb and electron–phonon interactions.
Integrating out the CDW collective fields $\phi_{\Gamma_7}$ and $\phi_{\Gamma_4}$ then yields the RPA renormalized phonon propagator.
\begin{eqnarray} && D^{\alpha\beta}_{\text{RPA}}(i \Omega_{m},\bm{q}) = \frac{1}{2} \Big\langle a_{\Gamma_{7}}^{\alpha}(i \Omega_{m},\bm{q}) a_{\Gamma_{7}}^{\beta \dagger}(i \Omega_{m},\bm{q}) \Big\rangle \nonumber \\ && = \Bigg\{ \Big(\Omega_{m}^{2} + \Omega_{\bm{q}}^{2}\Big) \delta_{\alpha\beta} - \frac{\lambda^{2}}{\Omega_{\bm{q}}} \Bigg(\frac{\delta_{\alpha\beta}}{V} \Big(\frac{\delta_{\beta\beta'}}{V} + \Pi^{\beta\beta'} \Big)^{-1} \Pi^{\beta'\beta} \nonumber \\ && -  \Pi^{\alpha\beta} \Bigg( \frac{\delta_{\beta\beta'}}{V} + \frac{\delta_{\beta\gamma}}{V} \Big(\frac{\delta_{\gamma\gamma'}}{V} + \Pi^{\gamma'\beta'} \Big)^{-1} \Pi^{\beta'\beta} \Bigg)^{-1} \Pi^{\beta'\beta}\Bigg) \Bigg\}^{-1}, \nonumber \\ \label{eq:ph_RPA} \end{eqnarray}

Eq.~(\ref{eq:ph_RPA}) provides the RPA renormalized phonon propagator used in the main text and demonstrates how the symmetry allowed $\Gamma_4$--$\Gamma_7$ mixing, amplified by electronic quantum fluctuations, drives phonon softening and the emergence of chiral CDW quantum criticality.

\section{Effective field theory for superconductivity}

In this section we derive the effective field theory describing the superconducting instability in the inter-orbital $\Gamma$–$L$ channel.
Our goal is to make explicit how the renormalized phonon and chiral CDW collective modes obtained within the RPA treatment enter the effective pairing interaction, and how they are combined with the Cooper pair susceptibility to determine the superconducting transition.

\subsection*{Total effective action}

We start from the total effective action
\begin{equation}
S_{\mathrm{eff}} = S_{\mathrm{el}} + S_{\mathrm{col}} + S_{\mathrm{vertex}},
\end{equation}
which combines the low-energy electronic degrees of freedom near the $\Gamma$ and $L$ points, the collective bosonic fields, and their mutual couplings.
The electronic part $S_{\mathrm{el}}$ follows from the low-energy Hamiltonians near the $\Gamma$ and $L$ points [Eq.~(\ref{eq:action_el})], while $S_{\mathrm{col}}$ encodes the dynamics of the $\Gamma_7$ phonon mode together with the $\Gamma_7$ and $\Gamma_4$ CDW collective fields [Eq.~(\ref{eq:collective})].

The coupling between electrons and collective bosonic modes is captured by the vertex action
\begin{eqnarray}
  && \mathcal{S}_{vertex}\Big(\psi_{\Gamma_{\alpha}},\psi_{L_{\alpha}},
  a_{\Gamma_{7}}^{\alpha},\phi_{\Gamma_{7}}^{\alpha},\phi_{\Gamma_{4}}^{\alpha}\Big)
  \nonumber \\
  && =  \frac{1}{\beta} \sum_{i \omega_{n}} \frac{1}{\beta} \sum_{i \Omega_{m}}
  \int \frac{d^{3} \bm{k}}{(2\pi)^{3}} \int \frac{d^{3} \bm{q}}{(2\pi)^{3}}
  \Big\{ \Big( \frac{\lambda}{\sqrt{\Omega_{\bm{q}}}} a_{\Gamma_{7}}^{\alpha}
  + \phi_{\Gamma_{7}}^{\alpha} + \phi_{\Gamma_{4}}^{\alpha} \Big)
  \nonumber \\
  && \hspace{2em} \times
  \psi_{\Gamma_{\alpha}}^{\dagger}(i \omega_{n} + i \Omega_{m},\bm{k} + \bm{q})
  \psi_{L_{\alpha}}(i \omega_{n},\bm{k}) + \mathrm{h.c.} \Big\} ,
  \label{eq:vertex}
\end{eqnarray}
which summarizes the electron–phonon and electron–CDW collective mode couplings relevant for pairing.

\subsection*{Integrating out collective modes}

We next integrate out all collective bosonic fields, including the $\Gamma_7$ phonon mode and both the $\Gamma_7$ and $\Gamma_4$ CDW collective fields.
Since these modes enter the action quadratically, this integration can be performed exactly at the Gaussian level.
The result is an effective electron–electron interaction characterized by an interaction kernel $V_{\mathrm{eff}}(i\Omega_m,\bm{q})$, which encodes the exchange of renormalized phonon and chiral CDW collective modes.
Within the same RPA framework used to describe the CDW sector, $V_{\mathrm{eff}}$ takes the form
\begin{eqnarray}
   && V_{eff}(i \Omega_{m},\bm{q}) = - \frac{\lambda^{2}}{{\Omega_{\bm{q}}}} D_{\alpha\alpha'}(i \Omega_{m},\bm{q}) + \mathcal{G}_{\phi, \alpha\alpha'}^{\Gamma_{7}}(i \Omega_{m},\bm{q}) \nn && + \frac{\lambda^{2}}{\Omega_{\bm{q}}} D_{\alpha\alpha'}(i \Omega_{m},\bm{q}) \Pi^{\alpha'\beta'}(i \Omega_{m},\bm{q}) \mathcal{G}_{\phi, \beta'\alpha'}^{\Gamma_{7}}(i \Omega_{m},\bm{q})\nn && + \frac{\lambda^{2}}{\Omega_{\bm{q}}} \mathcal{G}_{\phi, \alpha\gamma}^{\Gamma_{7}}(i \Omega_{m},\bm{q}) \Pi^{\gamma\gamma'}(i \Omega_{m},\bm{q}) D_{\gamma'\alpha'}(i \Omega_{m},\bm{q}) \nn && + \Big( \frac{\lambda^{2}}{\Omega_{\bm{q}}} \Big)^{2} D_{\alpha\alpha'}(i \Omega_{m},\bm{q}) \Pi^{\alpha'\beta'}(i \Omega_{m},\bm{q}) \mathcal{G}_{\phi, \beta'\gamma}^{\Gamma_{7}}(i \Omega_{m},\bm{q}) \nn && \times \Pi^{\gamma\gamma'}(i \Omega_{m},\bm{q}) D_{\gamma'\alpha'}(i \Omega_{m},\bm{q})  + \Big(\frac{\delta_{\alpha\alpha'}}{V} + \Pi^{\alpha\alpha'}(i \Omega_{m},\bm{q}) \Big)^{-1}, \nn
\end{eqnarray}
where $D_{\alpha\alpha'}(i \Omega_{m},\bm{q})$ is partially renormalized phonon mode,
\bqa && D_{\alpha\alpha'}(i \Omega_{m},\bm{q}) = \frac{1}{2} \Big\langle a_{\Gamma_{7}}^{\alpha}(i \Omega_{m},\bm{q}) a_{\Gamma_{7}}^{\alpha' \dagger}(i \Omega_{m},\bm{q}) \Big\rangle \nn && = \Bigg( \Big(\Omega_{m}^{2} + \Omega_{\bm{q}}^{2}\Big) \delta_{\alpha\alpha'} - \frac{\lambda^{2}}{\Omega_{\bm{q}}} \frac{\delta_{\alpha\beta}}{V} \Big(\frac{\delta_{\beta\beta'}}{V} + \Pi^{\beta\beta'} \Big)^{-1} \Pi^{\beta'\alpha'} \Bigg)^{-1}, \nn  \eqa
and $\mathcal{G}_{\phi, \alpha\alpha'}^{\Gamma_{7}}(i \Omega_{m},\bm{q})$ is partially renormalized CDW collective mode,
\bqa && \mathcal{G}_{\phi, \alpha\alpha'}^{\Gamma_{7}}(i \Omega_{m},\bm{q}) = \frac{1}{2} \Big\langle \phi_{\Gamma_{7}}^{\alpha}(i \Omega_{m},\bm{q}) \phi_{\Gamma_{7}}^{\alpha' \dagger}(i \Omega_{m},\bm{q}) \Big\rangle \nn && = \Bigg(\frac{\delta_{\alpha\alpha'}}{V} + \frac{\delta_{\alpha\beta}}{V} \Big(\frac{\delta_{\beta\beta'}}{V} + \Pi^{\beta\beta'} \Big)^{-1} \Pi^{\beta'\alpha'} \Bigg)^{-1} . \nn \eqa

\subsection*{Cooper channel and pairing susceptibility}

To describe superconductivity between the $p$-orbital band at $\Gamma$ and the $t_{2g}$ band at $L$, we introduce an inter-orbital Cooper pair field $\Delta$ in the $\Gamma$–$L$ channel via a Hubbard–Stratonovich transformation.
After introducing this field, we integrate out the fermionic degrees of freedom.
This yields a quadratic effective action for the Cooper pair field,
\begin{eqnarray}
  && \mathcal{S}_{\mathrm{eff}}
  \Big(\Delta_{\Gamma_{\alpha}L_{\alpha'}}(i\Omega_{m},\bm{q}) \Big)
  \nonumber \\
  && = \frac{1}{\beta} \sum_{i \Omega_{m}} \int \frac{d^{3} \bm{q}}{(2\pi)^{3}}
  \frac{1}{2}
  \Delta_{\Gamma_{\alpha}L_{\beta}}^{\dagger}(i\Omega_{m},\bm{q})
  \Big\{ (V_{\mathrm{eff}}(i\Omega_m,\bm{q}))^{-1}
  - \Pi_{\beta\beta'}^{pp}(i \Omega_{m},\bm{q}) \Big\}
  \Delta_{L_{\beta'}\Gamma_{\alpha'}}(i\Omega_{m},\bm{q}) . \nn
\end{eqnarray}

Here $\Pi^{pp}$ denotes the Cooper pair susceptibility in the particle–particle channel, constructed from the electronic Green functions at the $\Gamma$ and $L$ points as
\bqa && \Pi_{\alpha\alpha'}^{pp}(i \Omega_{m},\bm{q}) \nn && = \frac{1}{\beta} \sum_{i \omega_{n}} \int \frac{d^{3} \bm{k}}{(2\pi)^{3}} G^{\alpha\beta}_{\Gamma}(i \omega_{n}+i\Omega_{m},\bm{k}+\bm{q}) G^{\beta\alpha'}_{L}(-i \omega_{n},-\bm{k}) . \nn  \label{eq:pairsus_sm} \eqa

\section{Representation analysis for pairing symmetry}
In this section we document the symmetry classification procedure used for pairing structures in the six orbital basis. We provide the explicit $D_{3d}$ representation matrices acting in orbital space and we list the resulting basis functions for each irreducible representation relevant to $p$ to $d$ pairing. We also give the real space form factor analysis that connects the symmetry labels to the orbital selective pairing picture used in the main text.

The orbitals directly involved in the pairing are the three $p$-orbitals and the three $t_{2g}$ orbitals ($d_{xy}$, $d_{yz}$, $d_{zx}$).
In this six dimensional orbital space the $D_{\text{3d}}$ point group is represented by twelve $6\times6$ matrices namely the identity $E$, two $2\pi/3$ rotations $C_3$, three $\pi$ rotations $C_2'$, inversion $i$, two $iC_3$ operations and three $iC_2'$ operations.
For the even parity $t_{2g}$ orbitals the cyclic $C_3$ operation acts as $d_{xy} \rightarrow d_{yz} \rightarrow d_{zx}$ and $iC_3$ is equivalent to $C_3$.
For the odd parity $p$ orbitals $C_3$ again acts cyclically as $p_x \rightarrow p_y \rightarrow p_z$ while $iC_3$ introduces an additional sign change for the $p$ components for example $p_x \rightarrow -p_y$.
For the three $C'_2$ rotations acting on both $p$ and $t_{2g}$ orbitals the corresponding coordinate transformations are ($y \rightarrow -x$, $z \rightarrow -z$) ($x \rightarrow -z$, $y \rightarrow -y$) and ($z \rightarrow -y$, $x \rightarrow -x$).
Under inversion the $p$ orbitals change sign whereas the $t_{2g}$ orbitals remain unchanged.
Using these rules the symmetry generators U in the orbital basis ($p_x$, $p_y$, $p_z$, $d_{xy}$, $d_{yz}$, $d_{zx}$) take the following matrix forms.

\begin{eqnarray}
    &&U(I)=\begin{pmatrix}
        -1 & 0 & 0 & 0 & 0 & 0 \\
        0 & -1 & 0 & 0 & 0 & 0 \\
        0 & 0 & -1 & 0 & 0 & 0 \\
        0 & 0 & 0 & 1 & 0 & 0 \\
        0 & 0 & 0 & 0 & 1 & 0 \\
        0 & 0 & 0 & 0 & 0 & 1 \\
    \end{pmatrix},
\end{eqnarray}
\begin{eqnarray}
    &&U(C_3(1))=\begin{pmatrix}
        0 & 0 & 1 & 0 & 0 & 0 \\
        1 & 0 & 0 & 0 & 0 & 0 \\
        0 & 1 & 0 & 0 & 0 & 0 \\
        0 & 0 & 0 & 0 & 0 & 1 \\
        0 & 0 & 0 & 1 & 0 & 0 \\
        0 & 0 & 0 & 0 & 1 & 0 \\
    \end{pmatrix},
\end{eqnarray}
\begin{eqnarray}
    &&U(C_3(2)) = U(C_3(1))^T,
\end{eqnarray}
\begin{eqnarray}
    &&U(iC_3(1))=\begin{pmatrix}
        0 & 0 & -1 & 0 & 0 & 0 \\
        -1 & 0 & 0 & 0 & 0 & 0 \\
        0 & 1 & 0 & 0 & 0 & 0 \\
        0 & 0 & 0 & 0 & 0 & 1 \\
        0 & 0 & 0 & 1 & 0 & 0 \\
        0 & 0 & 0 & 0 & 1 & 0 \\
    \end{pmatrix},
\end{eqnarray}
\begin{eqnarray}
    &&U(iC_3(2)) = U(iC_3(1))^T,
\end{eqnarray}
\begin{eqnarray}
    &&U(C'_2(1))=\begin{pmatrix}
        0 & 1 & 0 & 0 & 0 & 0 \\
        \mp1 & 0 & 0 & 0 & 0 & 0 \\
        0 & 0 & \mp1 & 0 & 0 & 0 \\
        0 & 0 & 0 & 0 & 1 & 0 \\
        0 & 0 & 0 & -1 & 0 & 0 \\
        0 & 0 & 0 & 0 & 0 & -1 \\
    \end{pmatrix},
    \label{eq:UC21}
\end{eqnarray}
\begin{eqnarray}
    &&U(C'_2(2))=\begin{pmatrix}
        \mp1 & 0 & 0 & 0 & 0 & 0 \\
        0 & 0 & 1 & 0 & 0 & 0 \\
        0 & \mp1 & 0 & 0 & 0 & 0 \\
        0 & 0 & 0 & -1 & 0 & 0 \\
        0 & 0 & 0 & 0 & 0 & 1 \\
        0 & 0 & 0 & 0 & -1 & 0 \\
    \end{pmatrix},
\end{eqnarray}
\begin{eqnarray}
    &&U(C'_2(3))=\begin{pmatrix}
        0 & 0 & 1 & 0 & 0 & 0 \\
        0 & \mp1 & 0 & 0 & 0 & 0 \\
        \mp1 & 0 & 0 & 0 & 0 & 0 \\
        0 & 0 & 0 & 0 & 0 & 1 \\
        0 & 0 & 0 & 0 & -1 & 0 \\
        0 & 0 & 0 & -1 & 0 & 0 \\
    \end{pmatrix}.
    \label{eq:UC23}
\end{eqnarray}
In Eqs.(\ref{eq:UC21})-(\ref{eq:UC23}) the plus sign in the $p$ orbital block corresponds to the $iC'_2$ operations.
The matrix basis for the pairing functions in the same orbital space can also be constructed.\cite{SM_Kaba2019}
Excluding intraorbital pairing there are eighteen independent interorbital basis matrices for even and odd parity.

For example the $p_x–d_{xy}$ channel is represented by
\begin{eqnarray}
    \widehat{p_x d_{xy}} = \begin{pmatrix}
        0 & 0 & 0 & 1 & 0 & 0 \\
        0 & 0 & 0 & 0 & 0 & 0 \\
        0 & 0 & 0 & 0 & 0 & 0 \\
        \pm1 & 0 & 0 & 0 & 0 & 0\\
        0 & 0 & 0 & 0 & 0 & 0 \\
        0 & 0 & 0 & 0 & 0 & 0 \\
    \end{pmatrix},
\end{eqnarray}
where the plus and minus signs denote even and odd parity respectively.
The most general interorbital pairing function can then be written as
\begin{eqnarray}
    \hat{\Delta} = \sum_{i,j}f_{ij}(k_x,k_y,k_z)\widehat{p_i d_j},
\end{eqnarray}
where $i=x, y, z$ and $j = xy, yz, zx$.
Under a point group operation $g$ the pairing matrix transforms as
\begin{eqnarray}
    \Delta \rightarrow \Delta'=U(g)\Delta U(g)^T.
\end{eqnarray}
The explicit basis functions obtained from this representation analysis are summarized in Table \ref{table:pairing_a1u} and \ref{table:pairing_a2u} and complement the qualitative discussion in the main text.
We find pairing states that transform according to the $A_{1u}$ and $A_{2u}$ irreducible representations with various orbital combinations.
For instance the basis function $k_xk_y (\sum_{i=j}\widehat{p_id_j}-\sum_{i\neq j}\widehat{p_id_j})$ mixes all $p–d$ channels while giving the $i = j$ components $p_x–d_{xy}$, $p_{y}–d_{yz}$ and $p_z–d_{zx}$ the opposite sign relative to the other channels that share the same $k_xk_y$ form factor.
This structure shows that orbital selective pairing arises naturally in the symmetry classification.\\
We next validate these symmetry based gap functions using a simple tight-binding real space construction.
When electrons in the Ti $d$ orbitals at the origin form Cooper pairs with electrons in the $p$ orbitals on the six surrounding Se sites the real space pairing form factor is $ f_{r,r'} = \delta_{r,r'} $.

Here, \( r  = (0,0,0)\) labels the Ti site and
\begin{eqnarray}
    r'&&=\pm(a/2\sqrt{3},a/2,z_\text{Se}) \nn
    &&=\pm(a/2\sqrt{3},-a/2,z_\text{Se}) \nn
    &&=\pm(a/\sqrt{3},0,z_\text{Se})
\end{eqnarray}
denote the positions of the six Se neighbors where $a$ is the in plane lattice constant and $z_\text{Se}$ is the Ti–Se distance along the $z$ direction.\cite{SM_Kaneko2018}
The Fourier transform of this local form factor is
\begin{eqnarray}
    f_\mathbf{k}&&=\sum_{\mathbf{r-r'}}f_{\mathbf{r,r'}} e^{i\mathbf{k}(\mathbf{r-r'})}\nn&&=2\Big(\cos[\frac{a}{\sqrt{3}}k_x-z_\text{Se}k_z] \nn &&  +2\cos[\frac{a}{2}k_y] \cos[\frac{a}{2\sqrt{3}}k_x+z_\text{Se}k_z]\Big),\nn
    \label{eq:form}
\end{eqnarray}
for an even parity spatial wave function.
Expanding around $k = 0$ we obtain
\begin{eqnarray}\label{eq:tb_even}
    f_k^\text{even} \sim \text{const} + \alpha (k_x^2+ k_y^2) + \beta k_z^2 + \gamma k_x^2k_y^2k_z^2 \cdots
\end{eqnarray}
For odd parity states the cosine functions in Eq.(\ref{eq:form}) are replaced by sine functions together with an overall phase factor $i=e^{i\pi/2}$.
This yields
\begin{eqnarray}\label{eq:tb_odd}
    f_k^\text{odd} && \sim \alpha k_x + \beta k_z + \gamma k_xk_z(k_x+k_z) + \delta k_y^2(k_x+k_z) \cdots \nn
\end{eqnarray}
The coefficients in these expansions indicate which symmetry allowed basis functions can acquire large weight in a microscopic pairing state.
Although the detailed values depend on $a$ and $z_{Se}$ the dominant contribution comes from the constant term which corresponds to an essentially isotropic $s$ wave form factor.
Combined with the representation analysis this supports a $p–d$ pairing state whose leading component is a simple s wave with possible weak anisotropy proportional to $k_x^2+k_y^2+k_z^2$.
Basis functions with odd parity such as $k_x$ or $k_x^2k_z$ do not appear in Table.\ref{table:pairing_a1u} and \ref{table:pairing_a2u} for the interorbital $p–d$ pairing channel.
A notable outcome is that orbital selective $s$ wave pairing is symmetry allowed.
For example a combination of the form $\alpha\sum_{i=j}\widehat{p_id_j}-\beta\sum_{i\neq j}\widehat{p_id_j}$ implies that the $p_x-d_{xy}$ and $p_x-d_{yz}$ components can have opposite signs and different amplitudes.
Overall the analysis provides a complete set of symmetry allowed orbital selective $p–d$ pairing functions that can be directly compared with future spectroscopic measurements.

\begin{table}
\centering
\small
\begin{ruledtabular}
\begin{tabular}{cc}
Irrep & Pairing function \\
\hline
$A_{1u}$ &
Isotropic $s$-wave: $\alpha\sum_{i,j}\widehat{p_id_j}$ \\[2pt]
& Orbital selective isotropic $s$-wave:
$\alpha\sum_{i=j}\widehat{p_id_j}-\beta\sum_{i\neq j}\widehat{p_id_j}$ \\[2pt]
& Anisotropic $s$-wave:
$(k_x^2+k_y^2+k_z^2)\sum_{i,j}\widehat{p_id_j}$ \\[2pt]
& Orbital selective anisotropic $s$-wave: \\
& \quad $(k_x^2+k_y^2+k_z^2)\sum_{i=j}\widehat{p_id_j}$ \\
& \quad $(k_x^2+k_y^2+k_z^2)\sum_{i\neq j}\widehat{p_id_j}$ \\
& \quad $(k_x^2+k_y^2+k_z^2)(\sum_{i=j}\widehat{p_id_j}-\sum_{i\neq j}\widehat{p_id_j})$ \\[2pt]
& Possible pairing functions with nodes: \\
& \quad $k_xk_y\sum_{i,j}\widehat{p_id_j}$ \\
& \quad $k_xk_y\sum_{i=j}\widehat{p_id_j}$ \\
& \quad $k_xk_y(\sum_{i=j}\widehat{p_id_j}-\sum_{i\neq j}\widehat{p_id_j})$ \\
& \quad $k_xk_y\sum_{i=j}\widehat{p_id_j}\pm k_yk_z\sum_{i\neq j}\widehat{p_id_j}$ \\
& \quad $k_xk_y\sum_{i=j}\widehat{p_id_j}\pm k_zk_x\sum_{i\neq j}\widehat{p_id_j}$ \\
& \quad $(k_xk_y+k_yk_z+k_zk_x)\sum_{i,j}\widehat{p_id_j}$ \\
& \quad $(k_xk_y+k_yk_z+k_zk_x)\sum_{i=j}\widehat{p_id_j}$ \\
& \quad $(k_xk_y+k_yk_z+k_zk_x)\sum_{i\neq j}\widehat{p_id_j}$ \\
& \quad $(k_xk_y+k_yk_z+k_zk_x)(\sum_{i=j}\widehat{p_id_j}-\sum_{i\neq j}\widehat{p_id_j})$ \\
& \quad $k_xk_yk_z\sum_{i,j}\widehat{p_id_j}$ \\
& \quad $k_xk_yk_z\sum_{i\neq j}\widehat{p_id_j}$ \\
& \quad $k_xk_yk_z(\sum_{i=j}\widehat{p_id_j}-\sum_{i\neq j}\widehat{p_id_j})$ \\
\end{tabular}
\end{ruledtabular}
\caption{Interorbital pairing functions for the $A_{1u}$ representation.}
\label{table:pairing_a1u}
\end{table}

\begin{table}
\centering
\small
\begin{ruledtabular}
\begin{tabular}{cc}
Irrep & Pairing function \\
\hline
$A_{2u}$ &
Orbital selective isotropic $s$-wave:
$\alpha(\sum_{i}\widehat{p_id_{i+1}}-\sum_{i}\widehat{p_id_{i-1}})$ \\[2pt]
& Orbital selective anisotropic $s$-wave:
$(k_x^2+k_y^2+k_z^2)(\sum_{i}\widehat{p_id_{i+1}}-\sum_{i}\widehat{p_id_{i-1}})$ \\[2pt]
& Orbital selective pairing with nodes: \\
& \quad $(k_xk_y+k_yk_z+k_zk_x)(\sum_{i}\widehat{p_id_{i+1}}-\sum_{i}\widehat{p_id_{i-1}})$ \\
& \quad $k_xk_y(\sum_{i}\widehat{p_id_{i+1}}-\sum_{i}\widehat{p_id_{i-1}})$ \\
& \quad $k_yk_z(\sum_{i}\widehat{p_id_{i+1}}-\sum_{i}\widehat{p_id_{i-1}})$ \\
& \quad $k_zk_x(\sum_{i}\widehat{p_id_{i+1}}-\sum_{i}\widehat{p_id_{i-1}})$ \\
& \quad $k_xk_yk_z(\sum_{i}\widehat{p_id_{i+1}}-\sum_{i}\widehat{p_id_{i-1}})$ \\
\end{tabular}
\end{ruledtabular}
\caption{Interorbital pairing functions for the $A_{2u}$ representation.}
\label{table:pairing_a2u}
\end{table}

\begin{figure}[t]
  \centering
  \subfloat[]{\includegraphics[width=0.50\linewidth]{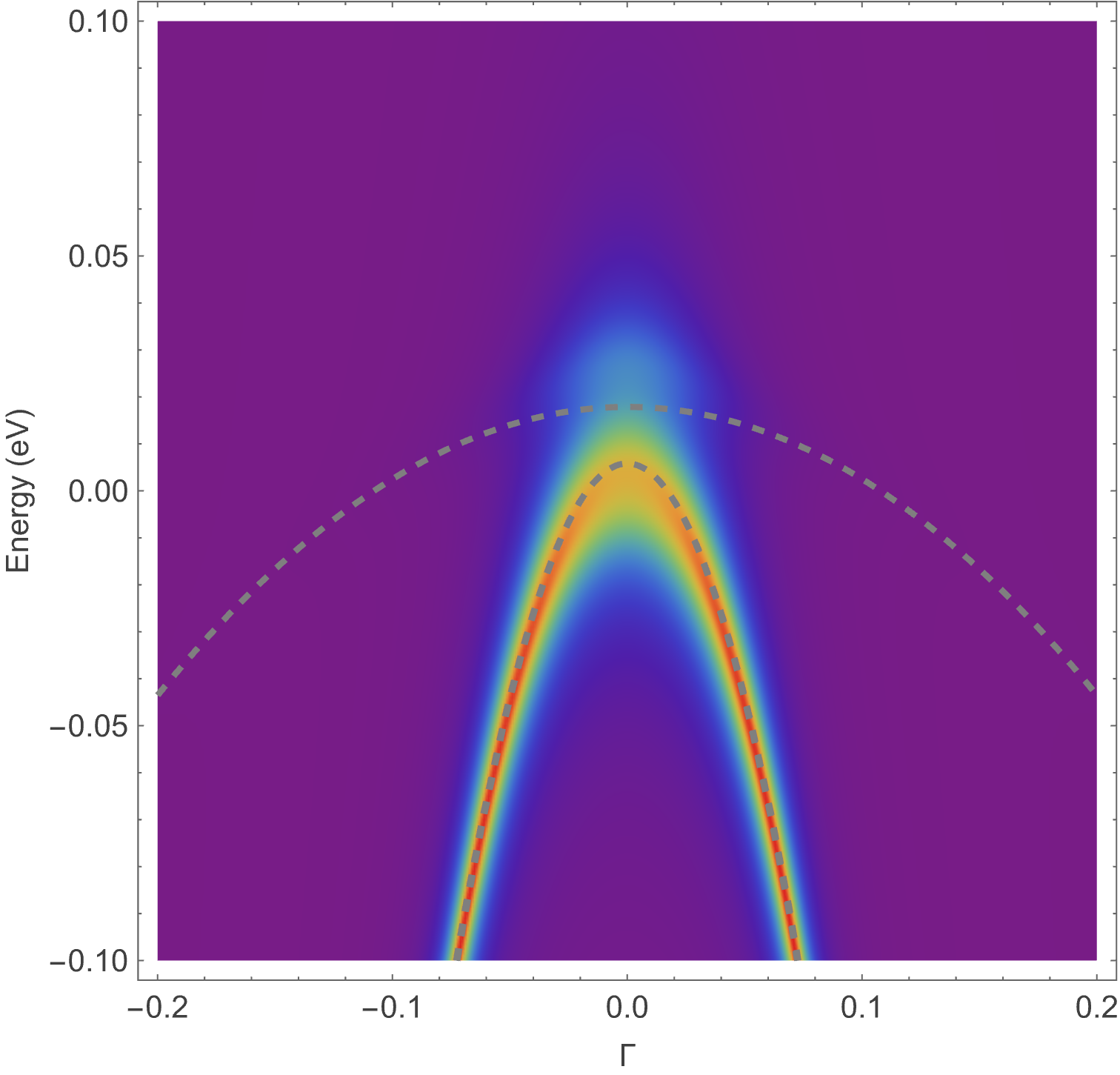}\label{fig:S1a}}
  \hfill
  \subfloat[]{\includegraphics[width=0.40\linewidth]{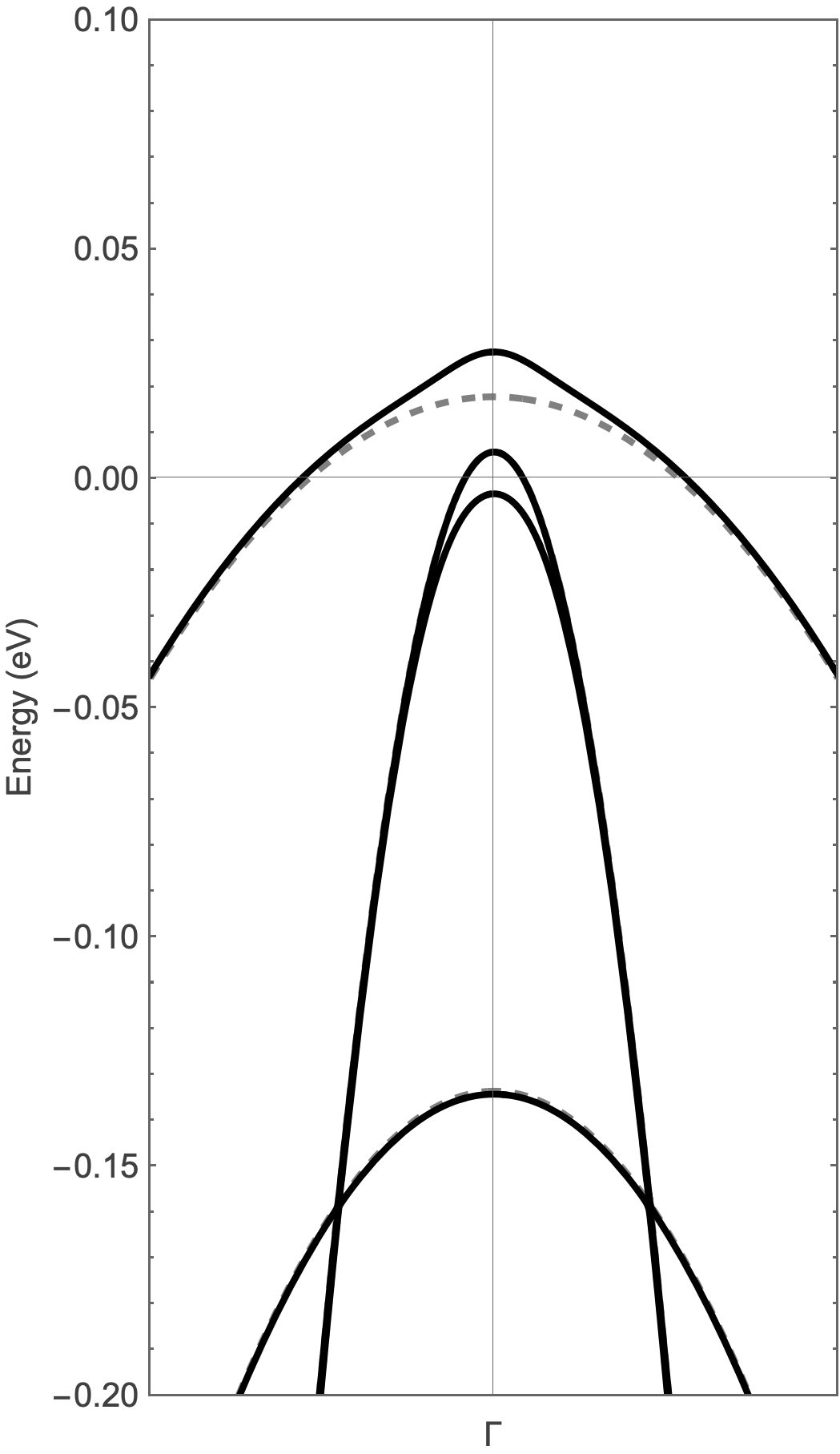}\label{fig:S1b}}
  \caption{Superconducting quasiparticle spectrum near the $\Gamma$ point.
(a) Spectral function $A(\mathbf{k},\omega)$ of the superconducting state evaluated near the $\Gamma$ point, calculated using the same method described in the Methods of the main text.
Similar to the behavior discussed in Fig.~4 for the $L$ point, the spectral intensity of the $\Gamma$-centered hole band is noticeably suppressed in the immediate vicinity of the $\Gamma$ point upon entering the superconducting state, reflecting the redistribution of spectral weight induced by inter-orbital finite-momentum pairing.
(b) Quasiparticle energies obtained by diagonalizing the effective Hamiltonian $H(\mathbf{k})$ defined in Eq.~(5) of the Methods; solid lines correspond to the superconducting state and dashed lines to the normal state.
Upon entering the superconducting state, the $L$-derived band folded to the $\Gamma$ point becomes renormalized, showing a slight upward bulging near the center, while the previously degenerate $\Gamma$-centered hole bands split with one branch shifting to lower energy.
}
  \label{fig:S1}
\end{figure}

\clearpage
\begin{SMbibliography}{60}

\bibitem{SM_Huang2021}Huang, S., Xu, S., Singh, B., Hsu, M., Hsu, C., Su, C., Bansil, A. \& Lin, H. Aspects of symmetry and topology in the charge density wave phase of 1T–TiSe$_2$. {\em New Journal Of Physics}. \textbf{23}, 083037 (2021), https://dx.doi.org/10.1088/1367-2630/ac1bf4

\bibitem{SM_Monney2011dft}Monney, C., Battaglia, C., Cercellier, H., Aebi, P. \& Beck, H. Exciton Condensation Driving the Periodic Lattice Distortion of 1T-TiSe$_2$. {\em Phys. Rev. Lett}. \textbf{106}, 106404 (2011), https://link.aps.org/doi/10.1103/PhysRevLett.106.106404

\bibitem{SM_Rasch2008}Rasch, J., Stemmler, T., Müller, B., Dudy, L. \& Manzke, R. 1T-TiSe$_2$: Semimetal or Semiconductor?. {\em Phys. Rev. Lett.}. \textbf{101}, 237602 (2008), https://link.aps.org/doi/10.1103/PhysRevLett.101.237602

\bibitem{SM_May2011}May, M., Brabetz, C., Janowitz, C. \& Manzke, R. Charge-Density-Wave Phase of 1T-TiSe$_2$: The Influence of Conduction Band Population. {\em Phys. Rev. Lett}. \textbf{107}, 176405 (2011), https://link.aps.org/doi/10.1103/PhysRevLett.107.176405

\bibitem{SM_Chen2015}Chen, P., Chan, Y., Fang, X., Zhang, Y., Chou, M., Mo, S., Hussain, Z., Fedorov, A. \& Chiang, T. Charge density wave transition in single-layer titanium diselenide. {\em Nature Communications}. \textbf{6}, 8943 (2015), https://doi.org/10.1038/ncomms9943

\bibitem{SM_Kim2024}Kim, K., Kim, H., Ha, S., Kim, H., Kim, J., Kim, J., Kwon, J., Seol, J., Jung, S., Kim, C., Ishikawa, D., Manjo, T., Fukui, H., Baron, A., Alatas, A., Said, A., Merz, M., Le Tacon, M., Bok, J., Kim, K. \& Kim, B. Origin of the chiral charge density wave in transition-metal dichalcogenide. {\em Nature Physics}. \textbf{20}, 1919-1926 (2024), https://doi.org/10.1038/s41567-024-02668-w

\bibitem{SM_Ishioka2010}Ishioka, J., Liu, Y., Shimatake, K., Kurosawa, T., Ichimura, K., Toda, Y., Oda, M. \& Tanda, S. Chiral Charge-Density Waves. {\em Phys. Rev. Lett.}. \textbf{105}, 176401 (2010), https://link.aps.org/doi/10.1103/PhysRevLett.105.176401

\bibitem{SM_Kaba2019}Kaba, S. \& Sénéchal, D. Group-theoretical classification of superconducting states of strontium ruthenate. {\em Phys. Rev. B}. \textbf{100}, 214507 (2019), https://link.aps.org/doi/10.1103/PhysRevB.100.21450

\bibitem{SM_Kaneko2018}Kaneko, T., Ohta, Y. \& Yunoki, S. Exciton-phonon cooperative mechanism of the triple-q charge-density-wave and antiferroelectric electron polarization in TiSe$_2$. {\em Phys. Rev. B}. \textbf{97}, 155131 (2018), https://link.aps.org/doi/10.1103/PhysRevB.97.155131

\end{SMbibliography}

\end{document}